\documentclass[review,authoryear,1p]{elsarticle}

\usepackage{setspace}
\usepackage{fullpage}
\usepackage{subfigure}
\usepackage{graphicx}
\usepackage{float}
\usepackage{amsmath,wasysym}
\usepackage[toc,page]{appendix}
\usepackage[small]{caption2}
\usepackage{booktabs}
\usepackage{widetext}
\usepackage{color,soul}
\def\bmath#1{\mbox{\boldmath$#1$}}
\newcommand{\ra}[1]{\renewcommand{\arraystretch}{#1}}

\usepackage{hyperref}
\hypersetup{
    colorlinks=true,
    linkcolor=blue,        
    citecolor=blue,  
    filecolor=magenta,
    urlcolor=blue,
    bookmarks=true,
}
\usepackage[all]{hypcap} 

\graphicspath{{/Figures}}

\renewcommand{\hl}{}

\journal{International Journal of Plasticity}
\begin{document}

\begin{frontmatter}
\title{An integrated full-field model of concurrent plastic deformation and microstructure evolution: Application to 3D simulation of dynamic recrystallization in polycrystalline copper}
\author[l1]{Pengyang Zhao}
\address[l1]{Department of Materials Science and Engineering, The Ohio State University}
\author[l1]{Thaddeus Song En Low}
\author[l1]{Yunzhi Wang\corref{cor1}}
\ead{wang.363@osu.edu}
\author[l1,l2]{Stephen R. Niezgoda\corref{cor1}}
\address[l2]{Department of Mechanical and Aerospace Engineering, The Ohio State University}
\ead{niezgoda.6@osu.edu}
\cortext[cor1]{Corresponding Author}

\begin{abstract}
Many time-dependent deformation processes at elevated temperatures produce significant concurrent microstructure changes that can alter the mechanical properties in a profound manner.
Such microstructure evolution is usually absent in mesoscale deformation models and simulations. Here we present an integrated full-field modeling scheme that couples the mechanical response with the underlying microstructure evolution. As a first demonstration, we integrate a fast Fourier transform-based elasto-viscoplastic (FFT-EVP) model with a phase-field (PF) recrystallization model, and carry out three-dimensional
simulations of dynamic recrystallization (DRX) in polycrystalline copper. A physics-based coupling
between FFT-EVP and PF is achieved by (1) adopting a
dislocation-based constitutive model in FFT-EVP, which allows the predicted dislocation density distribution to be converted to
a stored energy distribution and passed to PF, and (2) implementing a stochastic nucleation model
for DRX. Calibrated with the experimental DRX stress-strain
curves, the integrated model is able to deliver full-field mechanical and microstructural information,
from which quantitative
description and analysis of DRX can be achieved.
It is suggested that the initiation of DRX occurs significantly earlier than previous predictions, due to
heterogeneous deformation. DRX grains are revealed to form at both grain boundaries
and junctions (e.g., quadruple junctions) and tend to grow in a wedge-like fashion to maintain a
triple line (not necessarily in equilibrium) with old grains.
The resulting stress redistribution due to strain compatibility is
found to have a profound influence on the subsequent dislocation evolution and softening.
\end{abstract}

\end{frontmatter}

\section{Introduction}
Many thermomechanical processes and high temperature applications of materials involve
a coupled evolution of micromechanical fields, local defect populations, and microstructural constituents
including precipitates and grains.
For instance, during hot deformation crystalline materials with low stacking fault energy (SFE) often
undergo dynamic recrystallization (DRX) wherein new grains will
continue to nucleate and grow \citep{sakai1984overview,sakai2014dynamic}, altering the
population, mutual elastic interaction, and subsequent motion of dislocations in a distinctive manner. These microstructural
changes are difficult to capture through purely mechanical
laws \citep{roters2010overview}.The continued adoption of new computational
strategies for accelerated development of materials, such as Integrated Computational
Materials Engineering \citep{allison2006integrated,allison2011integrated},
for materials processed through thermomechanical routes
or materials exposed to thermal and mechanical extremes in-service requires modeling and
simulation tools that integrate both the mechanical and microstructural aspects in a fully coupled manner.

Within the broader solid mechanics and materials science communities,
mature computational approaches exist to study, separately, the deformation of heterogenous materials
and microstructure evolution, respectively.
On one hand, the field of crystal plasticity has benefited substantially
from multi-scale experiments and simulations linking mechanical properties
of materials with the evolution of
contained structural defects such as dislocations under an applied stress or
displacement field. At the mesoscale, this understanding has been implemented into
physics-based constitutive theories
\citep{arsenlis2002modeling,arsenlis2004evolution,cheong2004discrete,ma2006dislocation,gao2003geometrically,beyerlein2008dislocation}
and implemented into homogenized deformation models such as self-consistent schemes
\citep{lebensohn1993self,niezgoda2014stochastic} or
full-field simulations such as finite element based crystal plasticity (FE-CP)
\citep{kalidindi1992crystallographic,beaudoin1995hybrid,roters2010overview}
or fast Fourier transform (FFT) based crystal plasticity (FFT-CP) models
\citep{lebensohn2001n,lebensohn2012elasto,eisenlohr2013spectral}.
On the other hand, the microstructural evolution in crystals, such as grain growth \citep{chen1994computer,kazaryan2002grain,moelans2008quantitative}, static recrystallization \citep{moelans2013phase}, rafting in superalloy \citep{zhou2010large,gaubert2010coupling} and many other phenomena \citep{chen2002phase,wang2010phase} have been well studied using phase-field (PF) simulations. The non-boundary tracking field description of microstructures and
the incorporation of thermodynamics-based free energy formulation have made PF a very powerful
and robust tool in simulating and predicting the microstructural evolution often in a quantitative manner \citep{chen2002phase,boettinger2002phase,shen2004increasing,moelans2008quantitative,steinbach2009phase,wang2010phase}.

In recent years, efforts have been made towards developing CP models that can incorporate
microstructure features to study plastic deformation of materials such as austenitic steels,
TRIP steels, brass, TWIP steels and shape memory alloys that
deform not only by dislocation slips but also by
displacive phase transformation mechanisms. For instance, frameworks
\citep{thamburaja2001polycrystalline,turteltaub2005transformation,lan2005mesoscale,manchiraju2010coupling}
have been developed to incorporate martensitic transformations as the flow rules.
Mechanical twinning, which is of great importance to the plasticity of many BCC metals as well as
FCC metals with low SFE,
has also been incorporated into FE-CP models
\citep{kalidindi1998incorporation,staroselsky1998inelastic,salem2005strain,steinmetz2013revealing,zhang2008finite}.
\hl{Atomistically-informed dislocation-based models have also been developed recently and applied  to single crystal
\mbox{\citep{cereceda2015unraveling}}.}
In addition,
microstructure modeling techniques such as cellular automata and phase-field
have also been applied to the study of mechanics-induced microstructural
evolution such as static recrystallization
\citep{hesselbarth1991simulation,raabe2002cellular,takaki2007phase,moelans2013phase,chen2015integrated}.
These models,
while providing certain connection between microstructure and crystal plasticity, still lack
a dynamic coupling between the two. On the other hand,
phase-field models incorporating either continuum plasticity
\citep{gaubert2010coupling} or dislocation density fields \citep{zhou2010large} have also been
developed to study rafting in Ni-based superalloys. These models are mainly rooted in
the PF framework and significant advances are required in order to generalize the approach and incorporate a
wider range of existing constitutive theories.

Regarding the simulation of DRX, there have been models aiming to couple deformation with microstructure evolution. In the models of \cite{ding2001coupled} and \cite{takaki2008multi}, the growth of DRX grains was modeled by cellular automaton and PF, respectively, and the phenomenological Kocks-Mecking model \citep{mecking1981kinetics}
was used to model the evolution of average dislocation density, which in return influenced the mechanics
through its square root relationship with the flow stress.
Recently \cite{takaki2014multiscale} extended their previous work by replacing the flow stress model with an elastic-plastic FE, intending to establish a full coupling model. However, the mechanical behavior was still considered in a macroscopic level in the sense that an average dislocation density was assumed for each grain which evolved according to the Kocks-Mecking model \citep{mecking1981kinetics}. DRX, on the other hand, occurs at a sub-grain level where grain boundary (GB) bulging \citep{ponge1998necklace,wusatowska2002nucleation,miura2007nucleation}
or other possible mechanisms \citep{rollett2004recrystallization}
might operate to initiate the nucleation of new grains. This length scale separation implies that a
``homogenized'' coupling scheme \citep{takaki2014multiscale}
may limit its application in exploring the underlying physics.
Another recent DRX model by \cite{popova2014coupled}, which
couples FE-CP with a probabilistic cellular automata, successfully captured both texture and mechanical feature
during the DRX in magnesium. However, only geometrically necessary dislocations (GND) were considered with the effect of statistically stored dislocations (SSD) being ignored, and the two-dimensional (2D) simulation ignores important microstructural features
(e.g., the triple/quadruple grain junctions \citep{miura2005nucleation}) when analyzing the dynamics of DRX grains.

In this paper, we present an integrated modeling scheme of fully coupling the mechanical response
with the underlying microstructure evolution by employing FFT-CP and PF. This novel
scheme is then implemented in simulating dynamic recrystallization in polycrystalline copper in \hl{three-dimensions} (3D).
Quantitative agreement between simulations and experiments is achieved, and
the revealed underlying DRX structures and evolution, to the best of our knowledge,
are investigated in full-field 3D dynamics for the first time.

\section{Model}\label{sec:model}

\begin{figure*}[htb!]
\begin{center}
\subfigure[]{\label{fig:Model-AMS}\includegraphics[width=0.47\textwidth]{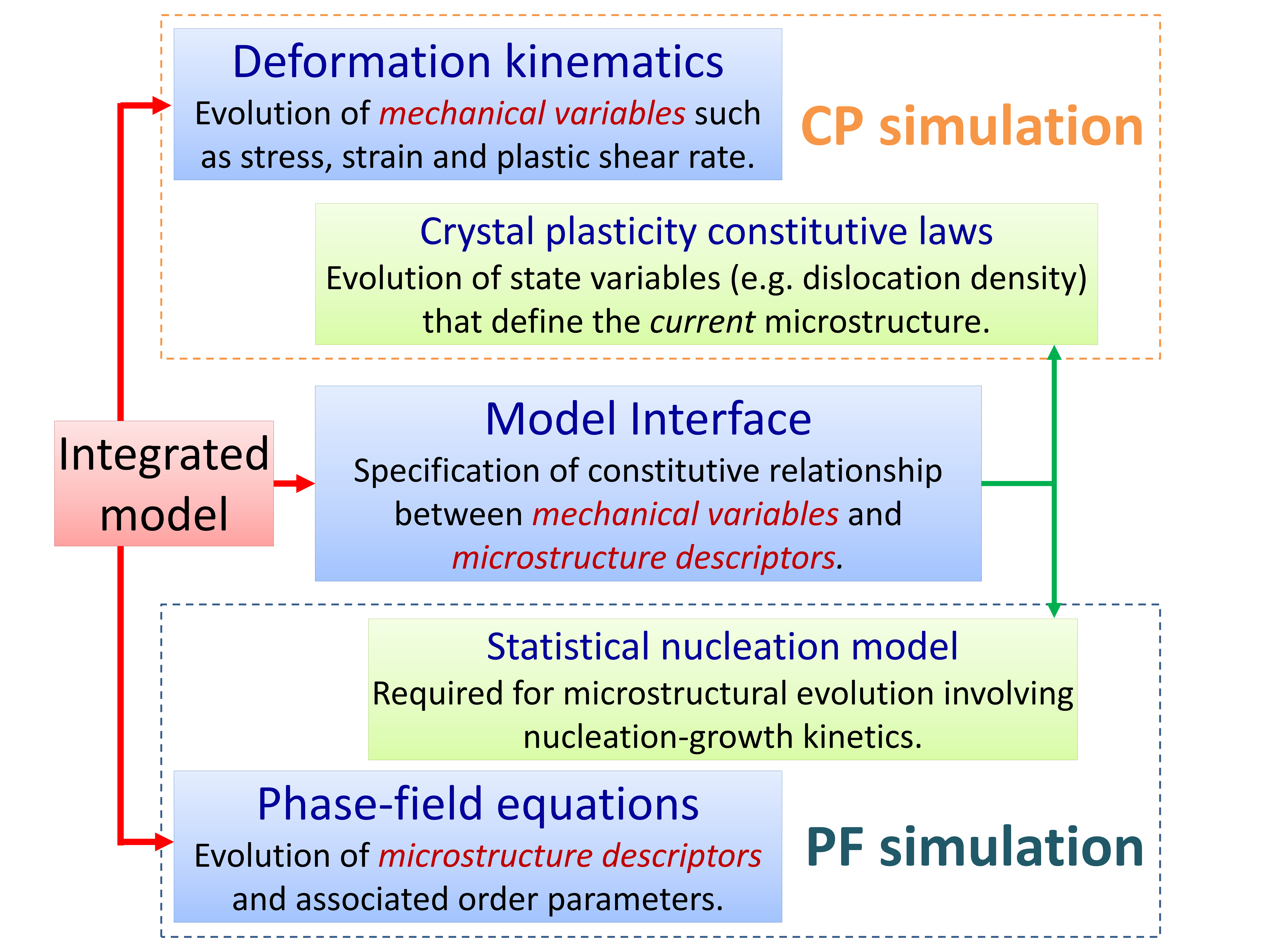}}
\hspace{10pt}
\subfigure[]{\label{fig:Model-CMS}\includegraphics[width=0.47\textwidth]{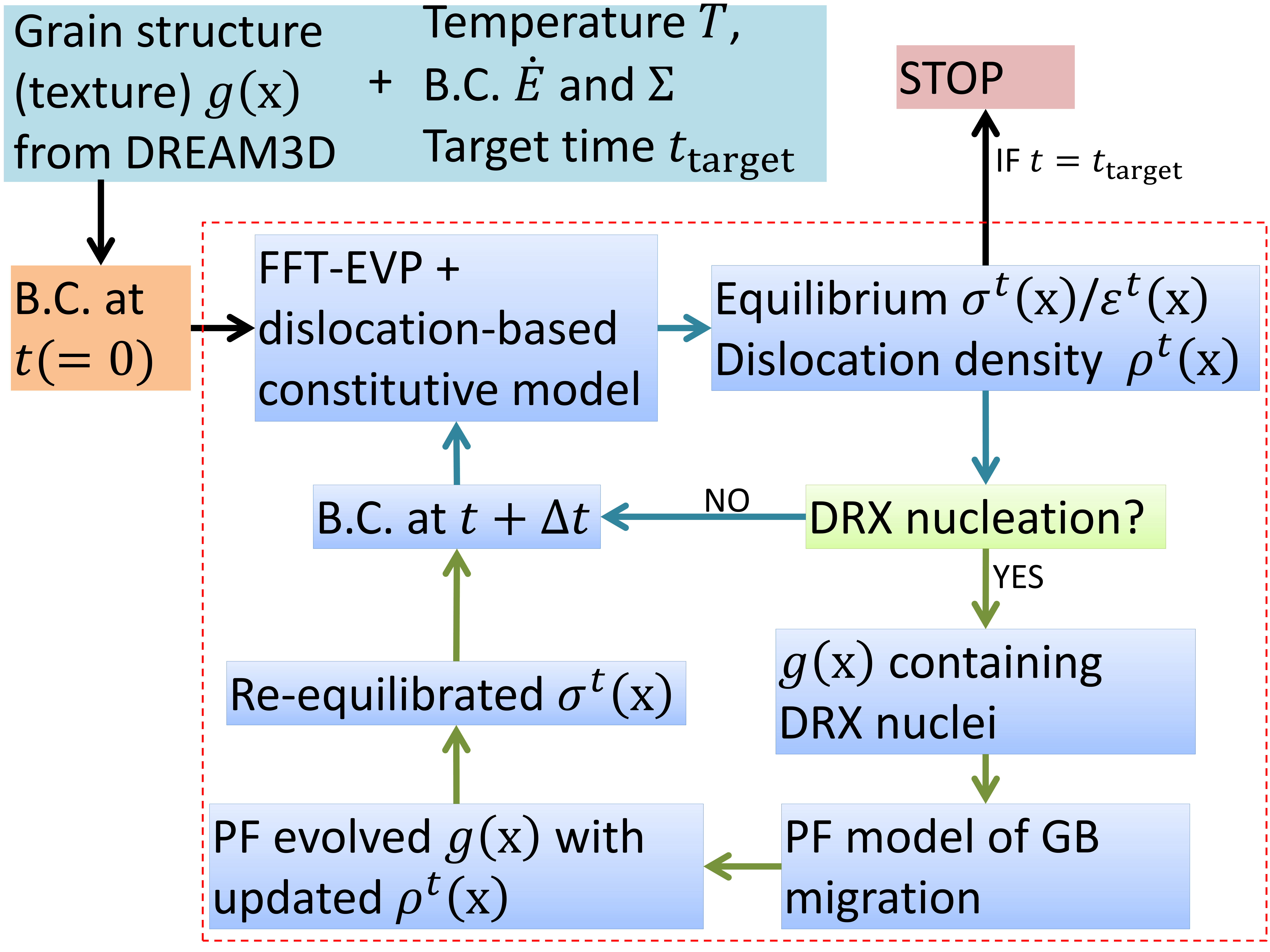}}
\caption{An integrated modeling scheme for thermal-mechanical processes: (a) a generic abstract model structure and
(b) a concrete model structure showing the flowchart for simulating dynamic recrystallization. Note that
once DRX is initiated, the nucleation occurs so frequently that
the PF microstructure relaxation occurs virtually every FFT-EVP step (see more detailed
discussion about implementing (b) in Sec. \ref{sec:UpdateMicrostructure}).}
\label{fig:Model-illus}
\end{center}
\end{figure*}

To better describe the integrated model that fully couples the mechanics and microstructure, we need to consider both the {\em abstract model structure} (AMS), which describes how the PF and mechanical simulation environments interact, and a {\em concrete model structure} (CMS), which describes the specific constitutive and kinetic models and what microstructural field variables are passed, as shown in Fig. \ref{fig:Model-illus}.
At the AMS level, the framework consists of three parts: (1) deformation kinematics,
(2) description of model interfaces, including a constitutive theory
that describes the evolution of a set of microstructural state variables (i.e., microscopic
variables such as dislocation density and arrangement, cell size, grain size, precipitate size,
and spacing, and so forth \citep{frost1982deformation}) and their interplay with PF order parameters,
and (3) PF governing equations describing
the kinetics of microstructure evolution as a function of order parameters. Additional nucleation models are needed when considering
issues involving the formation of new grains (e.g., DRX and twinning) or
new phases (e.g., precipitation) during the deformation.
Implementation of the AMS requires the identification of microstructure descriptors as well as CP frameworks,
constitutive laws, PF equations, and/or nucleation models, which all together describe the CMS.
As our first demonstration, we have chosen modeling DRX in
polycrystalline copper, of which the corresponding CMS with the basic simulation flowchart is shown in Fig. \ref{fig:Model-CMS}.
Based on the experiment of \cite{ghauri1990grain}, it was estimated that
during a compression test up to $130\%$ strain under applied strain rate of $1.6\times10^{-3}$ /s (conditions
to be used in the following simulation), grain growth in a polycrystal copper with an average grain size $>20$ $\mu$m
can be safely ignored (relative change $<2\%$). As a result, for efficiency and simplicity,
in Fig. \ref{fig:Model-CMS} PF is entered only when
DRX is initiated. In the following, the details of each model element will be presented.

\subsection{Deformation kinematics}\label{sec:FFT-EVP}
We employ the fast Fourier transform-based elasto-viscoplastic (FFT-EVP) formulation in the framework of infinitesimal-strain, developed by \cite{lebensohn2012elasto}. Using an Euler implicit time discretization and anisotropic elasticity, the stress in material point ${\bf x}$ at $t+\Delta t$ is given by
\begin{align}
\label{eq:HookeLaw}
\bmath{\sigma}^{t+\Delta t}\left({\bf x}\right)\;&=\;{\bf C}\left({\bf x}\right):\bmath{\varepsilon}^{{\rm e},t+\Delta t}(\mathbf{x})\; \\
&=\;{\bf C}\left({\bf x}\right):\left[\bmath{\varepsilon}^{t+\Delta t}\left({\bf x}\right)-\bmath{\varepsilon}^{{\rm p},t}-\dot{\bmath{\varepsilon}}^{{\rm p},t+\Delta t}\left({\bf x},\bmath{\sigma}^{t+\Delta t}\right)\Delta t\right] \nonumber
\end{align}
where $\bf C$ is the elastic stiffness tensor, $\bmath{\sigma}$ is the stress field, and the total strain field $\bmath{\varepsilon}$, in the small-strain assumption, is simply the sum of elastic and plastic part, i.e. $\bmath{\varepsilon}=\bmath{\varepsilon}^{\rm e}+\bmath{\varepsilon}^{\rm p}$. The plastic strain at $t +\Delta t$ 
is obtained by the first-order approximation $\bmath{\varepsilon}^{{\rm p},t+\Delta t}=\bmath{\varepsilon}^{{\rm p},t}+\dot{\bmath{\varepsilon}}^{{\rm p},t+\Delta t}\left(\bmath{\sigma}^{t+\Delta t}\right)\Delta t$. The usage of the plastic strain rate $\dot{\bmath{\varepsilon}}^{\rm p}$ at $t+\Delta t$ (a function of stress at $t+\Delta t$) in calculating the plastic shear increment from $t$ to $t+\Delta t$ suggests an implicit Euler method, which requires the stress field to be solved iteratively, here by a spectral approach which utilizes FFT algorithms to simplify convolution integrals involving Green's function. The plastic strain rate is formulated as
\begin{equation}
\dot{\bmath{\varepsilon}}^{\rm p}\left({\bf x}\right)\;=\;\sum_{\alpha=1}^\mathcal{N} {\bf m}^\alpha\left({\bf x}\right)\dot{\gamma}^\alpha\left({\bf x}\right)
\label{eq:PlasticRate}
\end{equation}
where $\dot{\gamma}^\alpha\left({\bf x}\right)$ and ${\bf m}^\alpha\left({\bf x}\right)$ are, respectively, the shear rate and the Schmid tensor associated with slip system $\alpha$ at point $\bf x$, and $\mathcal{N}$ is the number of active slip systems under consideration. The exact form of $\dot{\gamma}^\alpha$ is determined by the adopted constitutive laws (Sec. \ref{sec:Constitutive_Laws}), which are usually nonlinear equations and will be solved with an augmented Lagrangian scheme adapted from \cite{michel2000computational}. 

The small-strain assumption in Eq. (\ref{eq:HookeLaw}) implies that (i) the local neighborhood of a material point does not change drastically due to pure deformation, and (ii) the deformation increment is small with each numerical time step. These can be considered as legitimate in modeling DRX, of which the onset is still at a relatively small strain level (usually $\sim0.1$) \citep{sakai1984overview,rollett2004recrystallization}. In addition, once DRX is initiated, the mechanical behavior starts to be dominated by the migration of DRX grain boundaries modeled by PF, and the error due to small-strain assumption is further relieved. Additionally, the visco-plastic FFT (no elasticity) formulation with small-strain updates has been demonstrated by \cite{prakash2009simulation} to give comparable results to crystal-plasticity finite elements in the simulation of rolling in f.c.c. metals up to 60\% thickness reduction and drawing in b.c.c. metals up to 40\% equivalent strain. In that work it was found that even at large deformations the FFT scheme captured the location and distribution of stress fluctuations accurately, but that the specific numerical value of the stress localization varies with respect to the CP-FEM. These previous results provide confidence that stress/strain localizations predicted by the FFT-EVP will at least be statistically representative of those occurring in the material and sufficient to accurately predict evolution of the dislocation densities required for modeling nucleation and growth of new grains during DRX, even if the specific spatial values differ. Recently \cite{eisenlohr2013spectral} demonstrated a FFT formulation which incorporates finite deformation kinematics, which utilizes the 1st \hl{Piola-Kirchhoff} stress and the deformation gradient as work conjugate stress/deformation measures. Coupling of the finite strain FFT with phase field equations poses several additional computational difficulties, namely that the Fourier grid for the finite strain FFT is naturally formulated in the material or reference configuration while the spectral grid for the phase field is defined in the spatial or deformed state. Adoption of finite strain kinematics would require either a) explicitly mapping the material configuration to the spatial and interpolating onto a regular Fourier grid for the PF and performing the inverse mapping/interpolation every simulation time/strain increment or b) reformulating the phase field in the undeformed configuration. Another possible correction would be to adopt the approach of \cite{cuitino1992material} which uses a logarithmic mapping to extend the small-strain updates to the finite deformation regime. However that approach requires the approximation that the plastic spin is identically zero and the principal directions of the plastic strain are aligned with the principal stresses. For these reasons we adopt the general FFT-EVP framework of \cite{lebensohn2012elasto}, keeping in mind the inherent limitations. Integration of the finite strain FFT and phase field is ongoing and will be reported in future work. Specific discussion on the limitations and
implications of the small strain kinematics with respect to the specific DRX case study investigated can be found in Sec. \ref{sec:results}.  

\subsection{Crystal plasticity constitutive laws}\label{sec:Constitutive_Laws}
The model interface should provide a methodology link between the deformation mechanics and microstructure evolution such that the integrated modeling is physically self-consistent. This requires the shear rate in Eq. (\ref{eq:PlasticRate}) to be formulated based on a set of state variables that can also be used as microstructure descriptors.
Following the spirit of Orowan equation $\dot{\gamma}=\rho_{\rm M} b v$ where
$\rho_{\rm M}$ is the mobile dislocation density, $b$ the magnitude of Burgers vector, and $v$ the average dislocation velocity, 
\cite{ma2004constitutive} and \cite{ma2006dislocation} derived a dislocation-based flow rule for single-phase FCC metals:
\begin{equation}
\dot{\gamma}^\alpha = \left\{ 
  \begin{array}{l l}
    0, & \quad |\tau^\alpha|\leq\tau_{\rm pass}^\alpha,\\
    \dot{\gamma}_0^\alpha\exp\left[-\frac{Q_{\rm slip}}{k_{\rm B}T}\right]\sinh\left[\frac{|\tau^\alpha|-\tau_{\rm pass}^\alpha}{\tau_{\rm cut}^\alpha} \right]{\rm sign}\left(\tau^\alpha\right),& \quad |\tau^\alpha|>\tau_{\rm pass}^\alpha,
  \end{array}
  \right.
  \label{eq:FlowRule}
\end{equation}
where $\dot{\gamma}^\alpha$ is the plastic shear rate for slip system $\alpha$ (among the total $\mathcal{N}$
active slip systems) and the pre-exponential factor
$\dot{\gamma}_0^\alpha=\frac{2k_{\rm B}T}{c_1c_3\mu b^2}\nu_{\rm D}\sqrt{\rho_{\rm P}^\alpha}$,
with the attempt frequency $\nu_{\rm D}$ being in the order of Debye frequency, $k_{\rm B}$ the Boltzmann constant,
$T$ the temperature, $\mu$ the shear modulus, and $c_i$ fitting parameters.
$Q_{\rm slip}$ is the effective activation energy barrier for gliding through distributed obstacles consisting
of parallel and forest dislocations with a density of $\rho_{\rm P}^\alpha$ and $\rho_{\rm F}^\alpha$, respectively,
which result in a passing stress $\tau_{\rm pass}^\alpha=c_1\mu b\sqrt{\rho_{\rm P}^\alpha}$ and
a cutting stress $\tau_{\rm cut}^\alpha=\frac{Q_{\rm slip}}{c_2c_3b^2}\sqrt{\rho_{\rm F}^\alpha}$, respectively.
According to \cite{ma2004constitutive} and \cite{ma2006dislocation}, $\rho_{\rm P}^\alpha$ and $\rho_{\rm F}^\alpha$
are given as
\begin{equation}
\begin{aligned}
\rho_{\rm F}^\alpha\;=\;\sum_{\beta=1}^\mathcal{N}&\rho_{\rm SSD}^\beta |\cos\left(\tilde{\bf n}^\alpha,\tilde{\bf t}^\beta\right)|+|\rho_{\rm GNDs}^\beta \cos\left(\tilde{\bf n}^\alpha,\tilde{\bf d}^\beta\right)|+
|\rho_{\rm GNDet}^\beta \cos\left(\tilde{\bf n}^\alpha,\tilde{\bf t}^\beta\right)|\\
&\;+|\rho_{\rm GNDen}^\beta \cos\left(\tilde{\bf n}^\alpha,\tilde{\bf n}^\beta\right)|,
\end{aligned}
\label{eq:RhoForest}
\end{equation}

\begin{equation}
\begin{aligned}
\rho_{\rm P}^\alpha\;=\;\sum_{\beta=1}^\mathcal{N}&\rho_{\rm SSD}^\beta |\sin\left(\tilde{\bf n}^\alpha,\tilde{\bf t}^\beta\right)|+|\rho_{\rm GNDs}^\beta \sin\left(\tilde{\bf n}^\alpha,\tilde{\bf d}^\beta\right)|+
|\rho_{\rm GNDet}^\beta \sin\left(\tilde{\bf n}^\alpha,\tilde{\bf t}^\beta\right)|\\
&\;+|\rho_{\rm GNDen}^\beta \sin\left(\tilde{\bf n}^\alpha,\tilde{\bf n}^\beta\right)|,
\end{aligned}
\label{eq:RhoParallel}
\end{equation}
where $\tilde{\bf n}^\alpha$, $\tilde{\bf d}^\alpha$ and $\tilde{\bf t}^\alpha$ are, respectively, slip plane normal, slip direction and sense vector of slip system $\alpha$. $\rho_{\rm GNDs}^\alpha$, $\rho_{\rm GNDet}^\alpha$ and $\rho_{\rm GNDen}^\alpha$
are related to the total GND density $\rho_{\rm GND}^\alpha$ of slip system $alpha$ and will be defined in a moment.
The SSD density $\rho_{\rm SSD}^\alpha$ evolves as \citep{ma2006dislocation}
\begin{equation}
\dot{\rho}_{\rm SSD}^\alpha\;=\;c_4\sqrt{\rho_{\rm F}^\alpha}\dot{\gamma}^\alpha+c_6d_{\rm dipole}^\alpha\rho_{\rm M}^\alpha\dot{\gamma}^\alpha-c_5\rho_{\rm SSD}^\alpha\dot{\gamma}^\alpha
-c_7\exp\left[-\frac{Q_{\rm bulk}}{k_{\rm B}T}\right]\frac{|\sigma_{\rm vm}^\alpha|}{k_{\rm B}T}\left(\rho_{\rm SSD}^\alpha\right)^2\left(\dot{\gamma}_{\rm vm}\right)^{c_8},
\label{eq:SSDEvolution}
\end{equation}
which accounts for (1) the lock forming between mobile and forest dislocations, (2) the dipole formation between mobile dislocations, (3) the athermal annihilation of two parallel dislocations with anti-parallel Burgers vectors within a critical distance and (4) thermal annihilation by edge dislocation climb. Here $d_{\rm dipole}=\frac{\sqrt{3}\mu b}{16\pi(1-\nu)\tau^\alpha}$ is the critical distance for dipole formation (with $\nu$ being the Poisson's ratio), $Q_{\rm bulk}$ the activation energy for self-diffusion, $\sigma_{\rm vm}^\alpha$ the von Mises equivalent stress and $\dot{\gamma}_{\rm vm}$ the von Mises equivalent shear rate.

The GND density $\rho_{\rm GND}^\alpha$, based on the crystallographic relationship, is decomposed into
$\rho_{\rm GNDs}^\alpha$, $\rho_{\rm GNDet}^\alpha$ and $\rho_{\rm GNDen}^\alpha$ \citep{ma2006dislocation}.
The rate of
these effective GND densities satisfy the conservation law
$\left(\dot{\rho}_{\rm GND}^\alpha\right)^2\;=\;\left(\dot{\rho}_{\rm GNDs}^\alpha\right)^2+\left(\dot{\rho}_{\rm GNDet}^\alpha\right)^2+\left(\dot{\rho}_{\rm GNDen}^\alpha\right)^2$
and can be determined from the deformation gradient following the pioneering works done by \cite{nye1953some}, \cite{dai1997geometrically} and \cite{dai1997geometricallyPhD}, which in small strain approximation reduce to \citep{dai1997geometricallyPhD}
\begin{equation}
\begin{aligned}
\dot{\rho}_{\rm GNDs}^\alpha\;&=\;-\frac{1}{b}\nabla\dot{\gamma}^\alpha\cdot\tilde{\bf t}^\alpha,\\
\dot{\rho}_{\rm GNDet}^\alpha\;&=\;\frac{1}{b}\nabla\dot{\gamma}^\alpha\cdot\tilde{\bf d}^\alpha,\\
\dot{\rho}_{\rm GNDen}^\alpha\;&=\;0.
\end{aligned}
\label{eq:GNDEvolution}
\end{equation}
The gradient terms in Eq. (\ref{eq:GNDEvolution}) involve the physical size of simulation grid $l_0$, which
is in principle determined by the physical length scale of the representative volume element (RVE). \hl{It
is worth noting that when numerically evaluating these gradient terms, forward or backward difference will
be used for grid points at grain boundaries and central difference for points in the grain interior.}
Calculation of GND through Eq. (\ref{eq:GNDEvolution}), however, ignore the effect that the grain
boundary or interface, where the deformation gradient usually build, can also act as a source/sink to
emit/absorb GND \citep{sun2000observations}. Recently, the saturation of GND has been studied both
experimentally \citep{kysar2010experimental} and theoretically \citep{oztop2013length}. In the current model,
for simplicity, we account for this effect via a phenomenological treatment. In calculating the gradient in
Eq. (\ref{eq:GNDEvolution}), instead of $l_0$, we use an ``effective'' length $l_{\rm eff}$, updated as
\begin{equation}
\label{eq:L0evolution}
l_{\rm eff}^{t+\Delta t}\;=\;l_{\rm eff}^{t}\left(1+\frac{\varepsilon_{\rm vm}}{\xi}\right),
\end{equation} where $\varepsilon_{\rm vm}$ is the {\em current} average von Mises equivalent strain and $\xi$ ($\gg 1$)
is a fitting parameter that will give rise to a saturated GND density as the deformation becomes sufficiently large.

\subsection{Statistical model of dynamic recrystallization nucleation}\label{sec:Statistical_Nucleation}
\begin{figure*}[ht]
\begin{center}
\subfigure[]{\label{fig:CellLength}\includegraphics[width=0.35\textwidth]{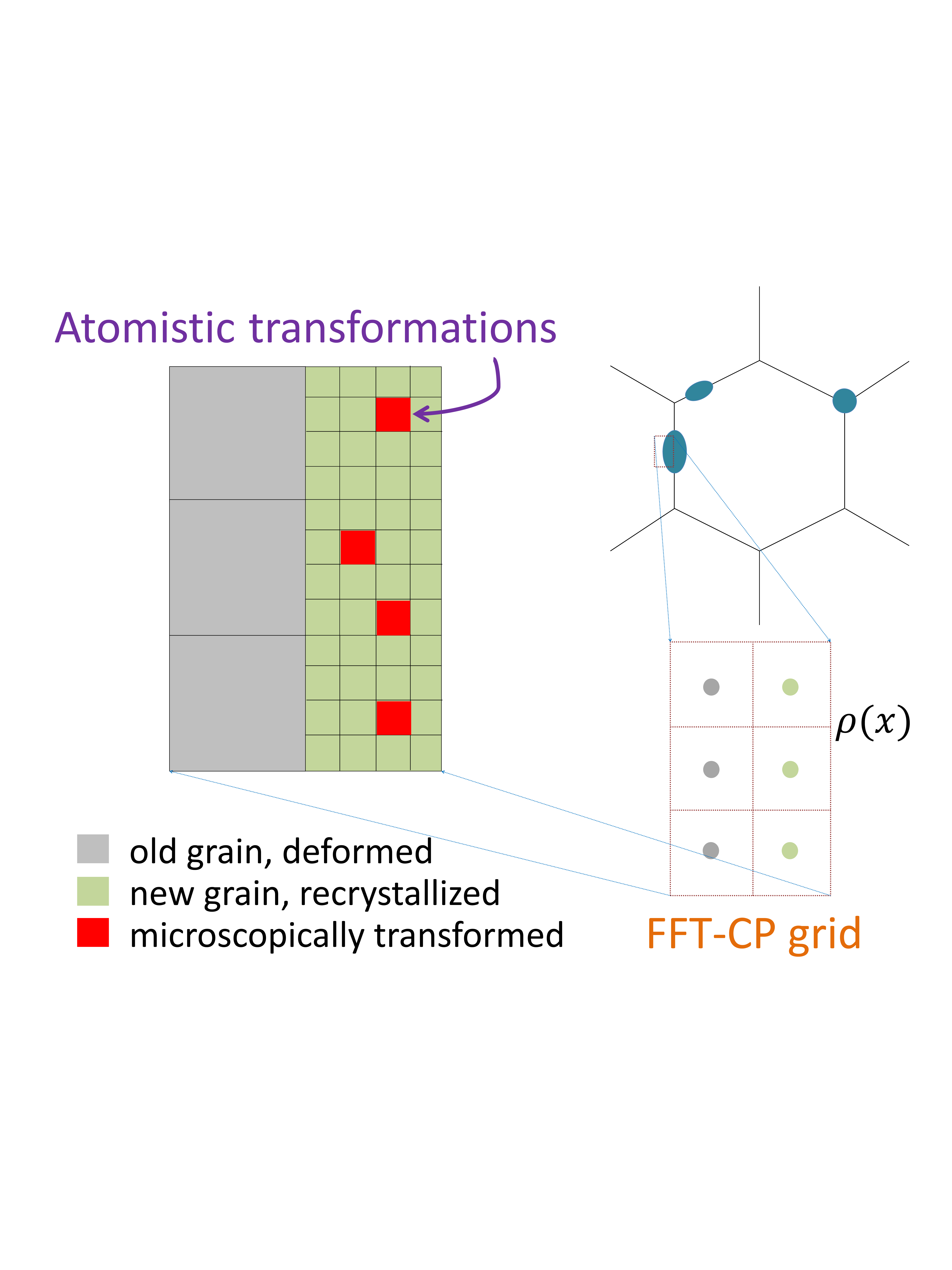}}
\subfigure[]{\label{fig:Prob-DRX-Nucl}\includegraphics[width=0.31\textwidth]{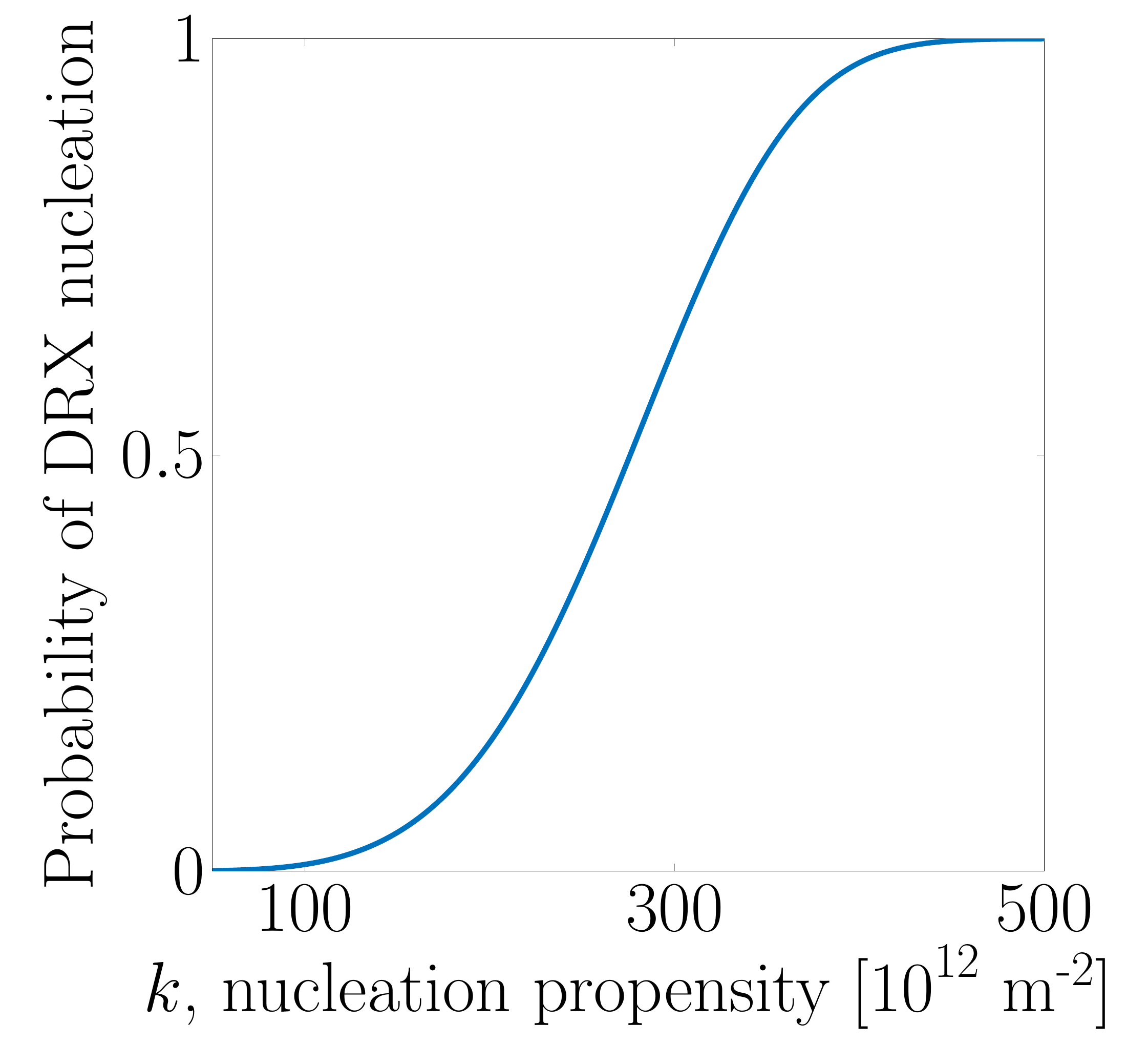}}
\hspace{0pt}
\subfigure[]{\label{fig:NuclStrength-prob}\includegraphics[width=0.31\textwidth]{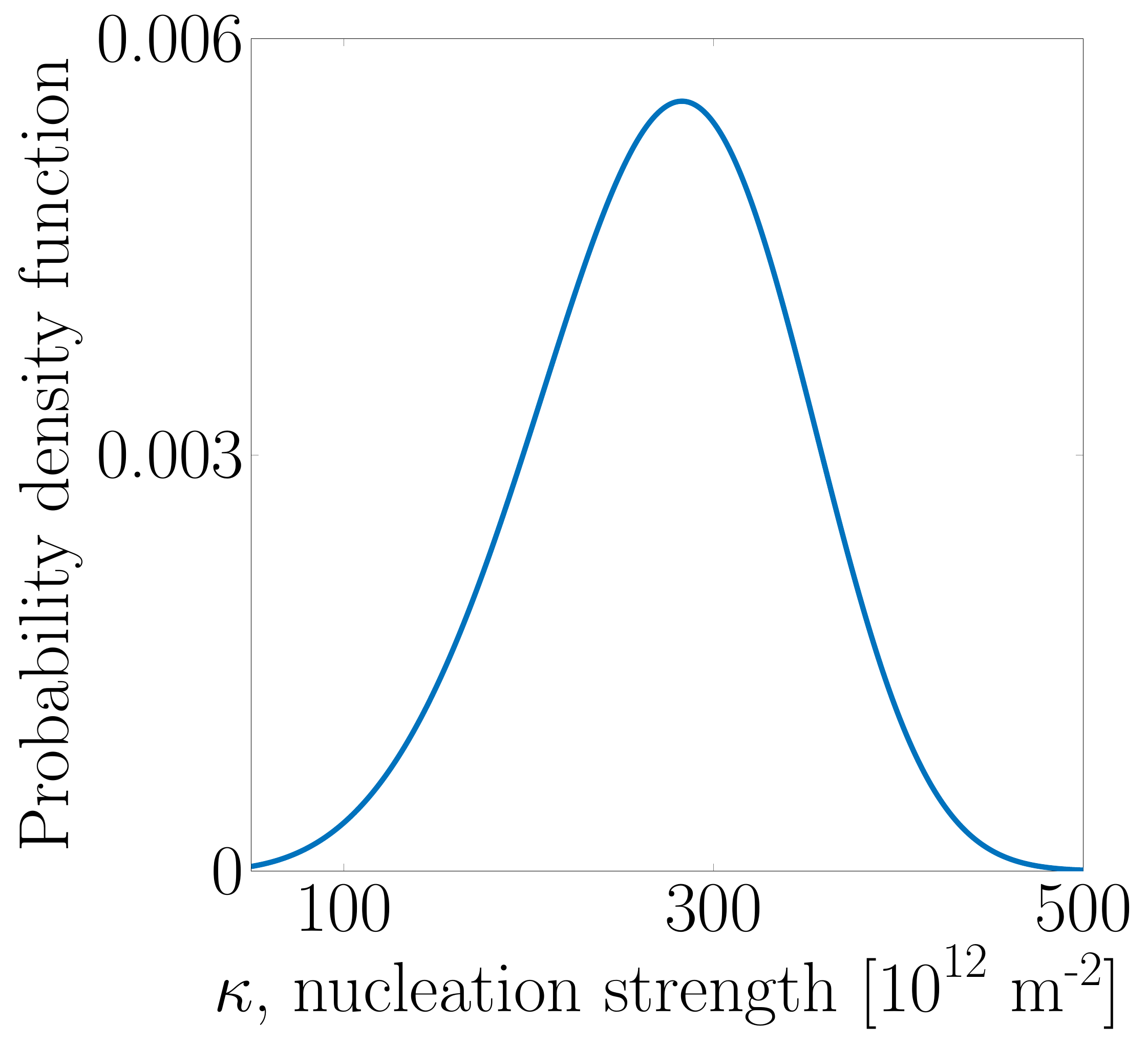}}
\caption{(a) Schematics of the incorporation of GB bulging mechanism in the current DRX nucleation model. Heterogeneous
microstructure is sampled using a regular FFT-CP grid, which can be divided into many sub-cells that can undergo discrete atomistic transformations. Based on our statistical modeling, the local
dislocation density $\rho({\bf x})$, defined on FFT-CP grid, is linked with
our predicted (b) probability of DRX nucleation and (c) the corresponding probability
density function of the nucleation strength.}
\label{fig:Illus-Nucl}
\end{center}
\end{figure*}

During dynamic recrystallization (DRX), new grains are initiated according to complicated and still
unclear atomistic mechanisms and usually correlated with stress/strain fields at continuum level
\citep{galiyev2001correlation,chauve2015strain}.
Together with considering the experimentally revealed grain boundary (GB) bulging mechanism \citep{ponge1998necklace,wusatowska2002nucleation,miura2007nucleation} , we make the following core assumptions: (a) DRX grains nucleate when some material at a GB undergoes certain {\em atomistic transformations} driven by thermal fluctuation and local stress to form a {\em stable} nucleus; (b) the time scale of this transformation and the formation of a stable nucleus is at atomic level and hence instantaneous as compared to the applied deformation, and (c) the length scale of this transformation is small relative to the FFT-CP computational grid. These consideration are similar to those made by 
\cite{niezgoda2014stochastic} in modeling of the stochastic twin nucleation, and is illustrated
in Fig. \ref{fig:CellLength} to show schematically the length scale separation.

To formulate a statistical model,
we assume that the underlying atomistic transformations, occurring on a finely divided subgrid at
each FFT-CP gridpoint \citep{simmons2000phase},
are independent and identically distributed and that a stable nucleus will form when some minimum number of subgrid points,
denoted $n^\ast$, transform.
Further we introduce a quantity $k({\bf x})$ at material point (FFT-CP grid) to measure
the {\em nucleation propensity} or the driving force for transformations, which,
as to be discussed below, will be a function of the local dislocation content that can be computed from FFT-CP.
Given that other factors that control nucleation are not explicitly included, we expect that the number of transformation events,
$\mathbf{N}$, is not strictly deterministic, but rather is a random discrete variable.
Our goal is then to link $\mathbf{N}$ and $k({\bf x})$ by formulating a continuous
distribution, which gives the probability of a stable nucleus forming at a specific GB (FFT-CP)
gridpoint given $k({\bf x})$, and then design a stochastic process accordingly for model implementation.
We leave the detail of derivation in \ref{appx:StatModelDerivation} and present directly the
probability of nucleating a DRX grain at a FFT-CP gridpoint (assuming $n^*=1$)
\begin{equation}
P\left({\bf N}(k)\geq 1\right)\;=\;1-e^{-\left(\frac{k}{k_c}\right)^q},
\label{eq:NumCrt}
\end{equation}
where $k_c=\left[C^{-1}\exp\left(\tfrac{Q_{\rm DRX}}{k_{\rm B}T}\right)\right]^{{1}/{q}}$ with $C$ and $q$ being
material constants and $Q_{\rm DRX}$ the apparent activation energy associated with DRX
\citep{ding2001coupled,rollett2004recrystallization}. To further appreciate Eq. (\ref{eq:NumCrt}), we introduce the concept of a stochastic  {\em nucleation strength} $\bmath{\kappa}({\bf x})$, which is a local material property
and depends on the subgrain/atomic structure. Nucleation will occur when the local propensity exceeds the local
strength, i.e., $k({\bf x})\;>\;\bmath{\kappa}({\bf x})$. The probability density of $\bmath{\kappa}$ can be determined as
\begin{equation}
f_{\bmath{\kappa}}(k)\;=\;\frac{dP\left(\bmath{\kappa}<k\right)}{dk}\;=\;\frac{dP\left({\bf N}(k)\geq n^*\right)}{dk}\;=\;\frac{q}{k_c}\left(\frac{k}{k_c}\right)^{q-1}e^{-\left(\frac{k}{k_c}\right)^q}.
\label{eq:NuclStrengthProb}
\end{equation}
The inherent stochastic nature we introduce here is the key distinctive feature compared with previous works \citep{ding2001coupled,takaki2008multi,takaki2014multiscale}.

The construction of the above statistical nucleation model is independent of the exact expression of
$k({\bf x})$. Here we simply use the total dislocation density
\begin{equation}
k({\bf x})\;\equiv\;\rho_{\rm tot}({\bf x})\;=\;\sum_{\alpha=1}^{\mathcal{N}}\rho_{\rm M}^\alpha({\bf x})+
\rho_{\rm SSD}^\alpha({\bf x})+\rho_{\rm GND}^\alpha({\bf x})
\label{eq:TotDislocationDensity}
\end{equation}
as the relevant measure of the DRX nucleation propensity. As the mechanistic details of nucleation are clarified, new forms
of $k({\bf x})$ can be incorporated without changing the overall framework. Figs. \ref{fig:Prob-DRX-Nucl} and
\ref{fig:NuclStrength-prob} plot, respectively, the probability of DRX nucleation
as a function of dislocation density (Eq. (\ref{eq:NumCrt})) and the probability density function (p.d.f.) of the nucleation strength
(Eq. (\ref{eq:NuclStrengthProb})), for the simulation to be described in Sec. \ref{sec:results}.
The parameters $k_c$ and $q$ were fit to the macroscale stress/strain curve shown in Fig. \ref{fig:SSs}.
\hl{Note that as nucleation only occurs at a small fraction of potential sites,} $k_c$ \hl{must be significantly larger than the average dislocation density. This relationship is explored in more detail in Sec.} \ref{sec:critical_strain}.

To implement the above statistical nucleation model, we initially draw a nucleation strength $\bmath{\kappa}({\bf x})$
for each FFT gridpoint ${\bf x}$ from the Weibull distribution (i.e., Eq. (\ref{eq:NumCrt})) specified by two parameters
$q$ and $k_c$.
Here $\bmath{\kappa}({\bf x})$ is considered as a static quantity during the deformation
until a DRX nucleation event occurs at ${\bf x}$, which will assign a new value to $\bmath{\kappa}({\bf x})$
due to the change of
atomistic environment. The reassignment of $\bmath{\kappa}({\bf x})$ again requires the specification of $q$ and $k_c$.
Recall that
$k_c=\left[C^{-1}\exp\left(\frac{Q_{\rm DRX}}{k_{\rm B}T}\right)\right]^{\frac{1}{q}}$, where $Q_{\rm DRX}$
is the activation energy characterizing the DRX nucleation mechanism and essentially related to
\hl{
the nature of the GB. Considering that the initial GB obtained via long term heat treatment
is apparently quite different from the boundary of recrystallized new grains, we use a different nucleation
parameter
}
$k_c^{'}$ to draw the new $\bmath{\kappa}({\bf x})$. Currently it is treated as $k_c^{'}=s_{\rm nucl}k_c$ with
$s_{\rm nucl}$ being simply a fitting parameter. In theory, $s_{\rm nucl}$ can
be completely determined once the knowledge of $Q_{\rm DRX}$, $q$ and $C$ are available.
\hl{
Since experiments have confirmed
that in copper the activation energy of migration of low angle GBs is much higher than that of high angle ones
} 
\citep{viswanathan1973kinetics},
it is expected that $s_{\rm nucl}$ could deviate significantly from 1 in reality. Nevertheless a further estimation of
$s_{\rm nucl}$ requires knowing the absolute value of nucleation rate (related to the coefficient $C$)
and is thus difficult to measure.

\subsection{Phase-field model}\label{sec:PF_Model}

The subsequent evolution of the grain structure containing nucleated DRX grains can be treated by extending the PF grain growth model of \cite{chen1994computer} by including the reduction of the stored strain energy, which competes with the reduction of the curvature to determine the migration of the recrystallized grain boundaries, as an additional driving force. \cite{moelans2013phase} have modified the PF grain growth model of
\cite{chen1994computer} by using two order parameters $\eta_{\rm rex}$ and $\eta_{\rm def}$ to describe one grain,
with the deformed state corresponding to $\left(\eta_{\rm rex}=0,\eta_{\rm def}=1\right)$ and the recrystallized state
to $\left(\eta_{\rm rex}=1,\eta_{\rm def}=0\right)$. We further extend the original 2D PF formulation of \cite{moelans2013phase} to 3D. The total free energy consists of GB energy and the stored strain energy, namely,
\begin{equation}
{F}_{\rm tot}\;=\;{F}_{\rm gb} + {F}_{\rm def}\;=\;\int f_{\rm gb}({\bf x})d{\bf x}+\int f_{\rm def}({\bf x})d{\bf x},
\label{eq:TotalFreeEnergy}
\end{equation}
where the local GB energy density $f_{\rm gb}$ (the dependence of ${\bf x}$ is omitted in what follows) is formulated as \citep{chen1994computer,moelans2013phase}
\begin{equation}
\begin{aligned}
f_{\rm gb}\;=\;&\sum_{i=1}^{N_g}\frac{6\omega_{\rm gb}}{l_{\rm gb}}\left[\frac{\eta_{{\rm rex},i}^4}{4}+\frac{\eta_{{\rm def},i}^4}{4}-\frac{\eta_{{\rm rex},i}^2}{2}-\frac{\eta_{{\rm def},i}^2}{2}+\frac{3}{2}\eta_{{\rm rex},i}^2\left(\sum_{j>i}^{N_g}\eta_{{\rm rex},j}^2+\sum_{j=1}^{N_g}\eta_{{\rm def},j}^2\right)\right.\\
&\left.+\frac{3}{2}\eta_{{\rm def},i}^2\left(\sum_{j=1}^{N_g}\eta_{{\rm rex},j}^2+\sum_{j>i}^{N_g}\eta_{{\rm def},j}^2\right)+\frac{1}{4}\right]+\frac{3}{8}\omega_{\rm gb}l_{\rm gb}\left(\left(\nabla\eta_{{\rm rex},i}\right)^2+\left(\nabla\eta_{{\rm def},i}\right)^2\right)
\end{aligned}
\label{eq:GBEnergyDensity}
\end{equation}
where $N_g$ is the total number of grain orientations under consideration, $\omega_{\rm gb}$ the GB energy
that can be taken from experiments, and $l_{\rm gb}$ the width of the diffuse interface zone associated with GB in PF method. The particular relationships between the coefficients were developed for the quantitative simulation of grain growth by \cite{moelans2008quantitative}. The stored energy density $f_{\rm def}$ is formulated as
\begin{equation}
f_{\rm def}\;=\;E_{\rm store}({\bf x})\frac{\sum_i^{N_g}\eta_{{\rm def},i}^2}{\sum_i^{N_g}\left(\eta_{{\rm rex},i}^2+\eta_{{\rm def},i}^2\right)}\;=\;\rho_{\rm tot}({\bf x})\zeta\mu b^2\frac{\sum_i^{N_g}\eta_{{\rm def},i}^2}{\sum_i^{N_g}\left(\eta_{{\rm rex},i}^2+\eta_{{\rm def},i}^2\right)}
\label{eq:StoredEnergyDensity}
\end{equation}
where the predicted stored energy $E_{\rm store}({\bf x})$ by FFT-EVP is approximated \citep{hull2001introduction} as $E_{\rm store}({\bf x})\approx\rho_{\rm tot}({\bf x})\zeta\mu b^2$ where $\zeta$ is a dimensionless constant of 0-1 and the total dislocation density is given in Eq. (\ref{eq:TotDislocationDensity}). The evolution equations of the order parameters simply follow the Allen-Cahn equation \citep{allen1979microscopic}, which in our case takes the following form \citep{moelans2008quantitative}:
\begin{equation}
\begin{aligned}
\frac{\partial \eta_{{\rm rex},i}}{\partial t}\;&=\;-\frac{4}{3}\left(\frac{M_{\rm gb}}{l_{\rm gb}}\right)\frac{\delta F_{\rm tot}}{\delta\eta_{{\rm rex},i}},\quad i=1\ldots N_g, \\
\frac{\partial \eta_{{\rm def},i}}{\partial t}\;&=\;-\frac{4}{3}\left(\frac{M_{\rm gb}}{l_{\rm gb}}\right)\frac{\delta F_{\rm tot}}{\delta\eta_{{\rm def},i}},\quad i=1\ldots N_g,
\end{aligned}
\label{eq:PFGoverning}
\end{equation}
where $M_{\rm gb}$ is the GB mobility. To model the property of a polycrystal, the RVE should contain sufficient number of grains, which leads to a large number of coupled partial differential equations in Eq. (\ref{eq:PFGoverning}).
To solve this numerical difficulty, we adopt the sparse data
structure technique proposed by \cite{gruber2006sparse} and \cite{vedantam2006efficient}
to reduce the effective $N_g$ considered for each PF node.
It also needs to be pointed out that the PF simulation does not need to reside on the same computational grid
as in FFT-EVP, owing to the different length scales of the corresponding physical processes as indicated in
Fig. \ref{fig:CellLength}. A finer PF grid can be used by interpolating the FFT-EVP grid to increase spatial resolution
when simulating the DRX grain growth. \hl{To do this, linear interpolation is used to obtain the finely sampled smooth fields such as
stress, strain, and dislocation densities; the refined grain structure with a refinement ratio
PF grid/FFT-EVP grid = 2:1 is obtained by assigning the orientation value of the
nearest original grid point to the new interpolating points, which is only involved in grain boundaries and junctions.}
\hl{The grain ID after PF simulation is determined by
\mbox{${\rm max}\left\lbrace\{\eta_{{\rm rex},i}, \eta_{{\rm def},i}\}_i^{N_g}\right\rbrace$}}.

If a time increment $\Delta t$ is used to solve the
deformation kinematics (Eq. (\ref{eq:HookeLaw})), the
corresponding number of PF steps $\Delta N_{\rm pf}$ for microstructure relaxation that can take place before
further mechanical loading is simply given as
$\Delta N_{\rm pf}\;=\;\lfloor\frac{\Delta t}{\delta t}\rfloor$
where $\delta t$ is the real time of one PF simulation step,
and $\lfloor x\rfloor$ means the largest integer not greater than $x$.
The determination of $\delta t$
follows \cite{moelans2008quantitativePRB} where parameters were taken so that 6[m$^3$/J]$\omega_{\rm gb}
\left[{\rm J}/{\rm m}^2\right]=l_{\rm gb}\left[{\rm m}\right]$,
$l_{\rm gb}=\sqrt{9.6}\delta x$, and $M_{\rm gb}\left[{\rm m}^4/({\rm J}\cdot{\rm s})\right]\delta t\left[{\rm s}\right]
=$0.0375[m$^3$/J]$l_{\rm gb}\left[{\rm m}\right]$ where
$\delta x$ is the discrete grid spacing in PF simulation.
Such numerical specification has been shown to yield a relative (numerical) error smaller than $5\%$
in terms of simulating GB motion \citep{moelans2008quantitativePRB}.
It is then given that
$\delta t\left[{\rm s}\right]\;=\;\frac{0.225\left[{\rm m}^6/{\rm J}^2\right]\omega_{\rm gb}\left[{\rm J}/{\rm m}^2\right]}{M_{\rm gb}\left[{\rm m}^4/({\rm J}\cdot{\rm s})\right]}$, which
as expected is controlled by GB energy and mobility.

In principle, GB energy and mobility should depend on the
misorientation and inclination, which leads to grain growth in anisotropic systems and has been considered
by previous PF models \citep{kazaryan2000generalized,ma2004computer,suwa2007three,moelans2008quantitativePRB}.
Nevertheless, the experiment \citep{wusatowska2002nucleation} to be simulated and compared shows mostly
equiaxed DRX grains, unlike what was predicted by PF grain growth in anisotropic system \citep{kazaryan2002grain};
the essentially random texture found in the experiment \citep{wusatowska2002nucleation} also frees us from
considering the high anisotropy in GB mobility that manifests only in
highly textured polycrystals \citep{ma2004computer,suwa2007three}. As a result, we assume isotropic grain growth
in the current model and take GB energy $\omega_{\rm gb}$ and mobility $M_{\rm gb}$ as constant.

\subsection{Microstructure update and stress redistribution}\label{sec:UpdateMicrostructure}
Once new grains are formed,
state variables, e.g., dislocation
density and grain orientation, should be updated based on
DRX mechanisms.
The general picture that DRX grains are ``dislocation-free'' or with a much lower dislocation
content describes the dislocation activities occurring at subgrain-level \citep{sakai1990dislocation}
and consequently cannot be directly applied to the dislocation density in the current model, which
is defined at a level much larger than the subgrains (Fig. \ref{fig:CellLength}). The
``dislocation-free'' nucleus at subgrain-level will co-deform with its surrounding grains and will rapidly accumulate a significant dislocation density before it grows to the length scale of
a typical continuum plasticity simulation.
This has been confirmed experimentally 
by transmission electron microscopy (TEM) observations \citep{sakai1990dislocation,sakai1995dynamic}.
Based on this length scale separation, we update the 
the GND density of material points in new grains as
\begin{equation}
\rho_{\rm GND}^{t+\Delta t}\;\leftarrow\;s_{\rm soften}\rho_{\rm GND}^{t}
\label{eq:UpdateGND}
\end{equation}
where $0<s_{\rm soften}<1$ is a fitting parameter, while keeping the SSD density unchanged.
Physically this means that we ignore the very initiation stage of
the growth of subgrain nuclei, which
has actually been implied in our nucleation model (Fig. \ref{fig:CellLength}). Numerically
this ensures the numerical stability in an otherwise high-contrast ``mixture'' of soft and hard grains.
The change of GND accounts for (i) the underlying change of subgrains orientation, i.e., dense dislocation
cell walls transforming into high angle boundaries, and (ii) the elimination of pre-existing GBs.
On the other hand, the SSD is expect to
be more or less the same at the coarse-grained FFT grid.

Regarding the crystal orientation, experiments have shown formation of moderate-angle
subboundaries or twins associated with DRX grains. However, there is no overall correlation
between DRX and the resulted grain orientations and the texture remains unchanged (or slightly more randomized)
during DRX \citep{wusatowska2002nucleation}.
For our first demonstrative case-study,
we simply assume that the crystal orientation of the new DRX grain points will inherit the current local values, leaving
the influence of DRX grain orientation to future studies.

Once the nucleation step is complete and the state variables updated,
subsequent PF relaxation will drive the new grains to consume the surrounding old grains. The change in local dislocation density and lattice orientation \hl{(the part due to deformation)} are accompanied by changes in the elastic strain fields and the stress needs to be re-calculated before the next deformation increment. If a new equilibrium stress state is not calculated, the resolved stresses in the now softer recrystallized grains will be significantly higher than the slip resistances, resulting in non-physical high strain rates.
\hl{In order to maintain the
compatibility, the previous strainrate field (before DRX) is prescribed and new equilibrium stress field due to the change of
microstructure is calculated before further mechanical loading is applied in FFT-EVP. The
deformation continues following the dynamically coupled scheme described above}
until the prescribed target strain increment is achieved, as shown in Fig. \ref{fig:Model-CMS}. It needs to be pointed out that once DRX has been initiated, the nucleation event occurs
so frequently that the PF microstructure relaxation occurs virtually every FFT-EVP step. As a result,
during the implementation of Fig. \ref{fig:Model-CMS}, DRX grains that have been grown for
a few steps are still allowed to grow as long as there is a driving force due to the store energy difference.

\section{Results}\label{sec:results}
\begin{figure*}[htb!]
\begin{center}
\subfigure[]{\label{fig:SSs}\includegraphics[width=0.38\textwidth]{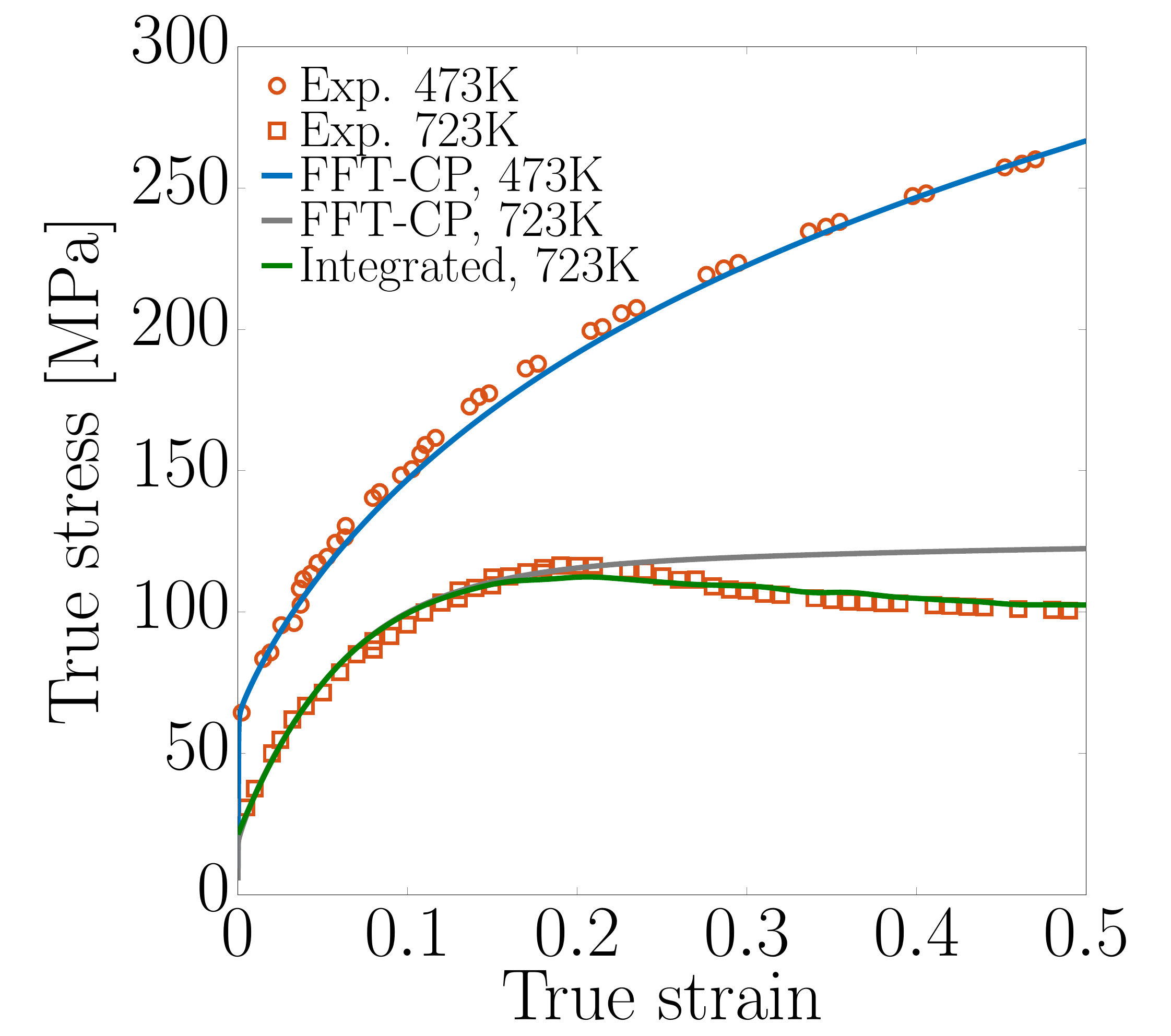}}
\hspace{14pt}
\subfigure[]{\label{fig:HRs}\includegraphics[width=0.38\textwidth]{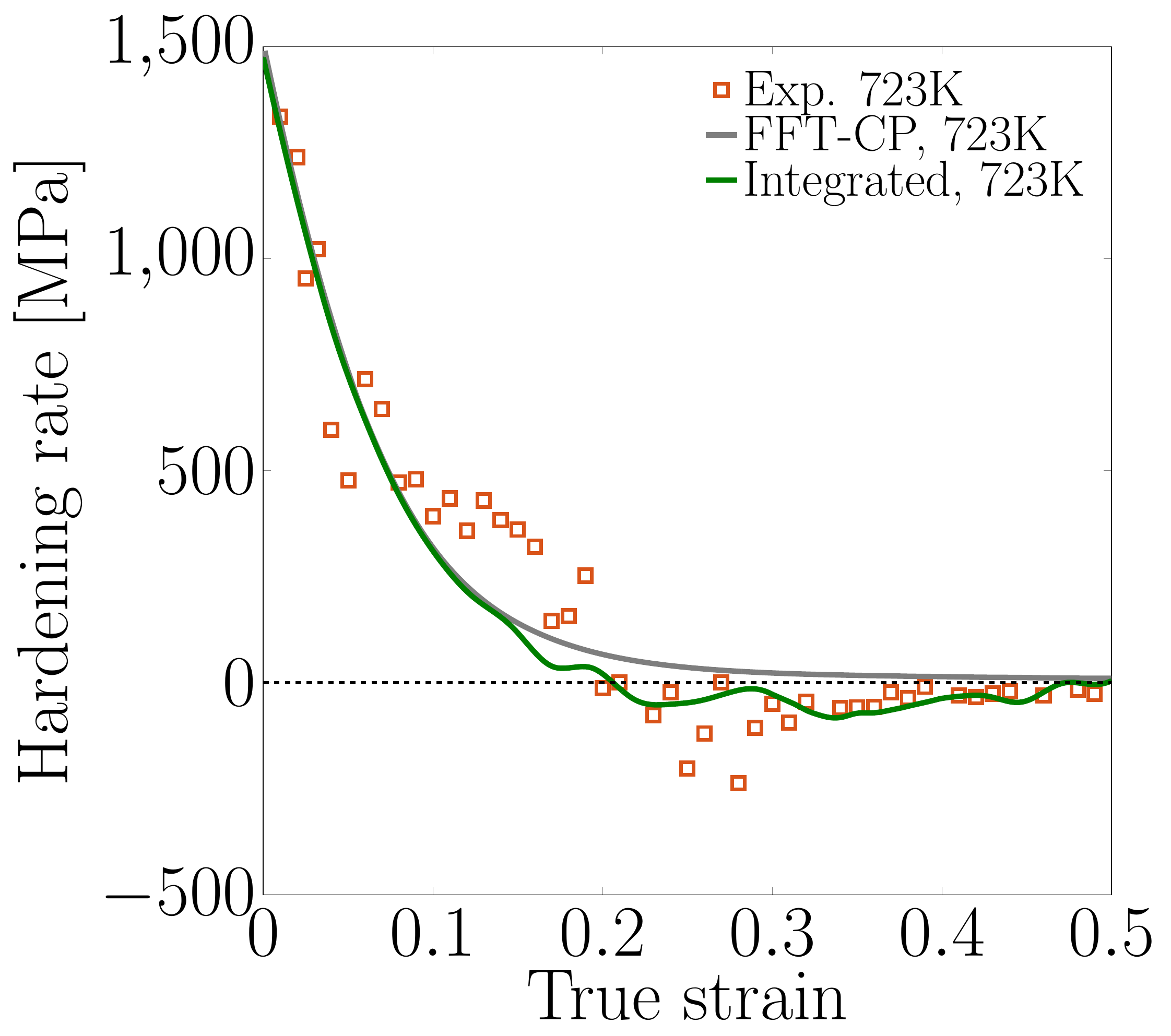}}\\
\subfigure[]{\label{fig:DD-473K}\includegraphics[width=0.38\textwidth]{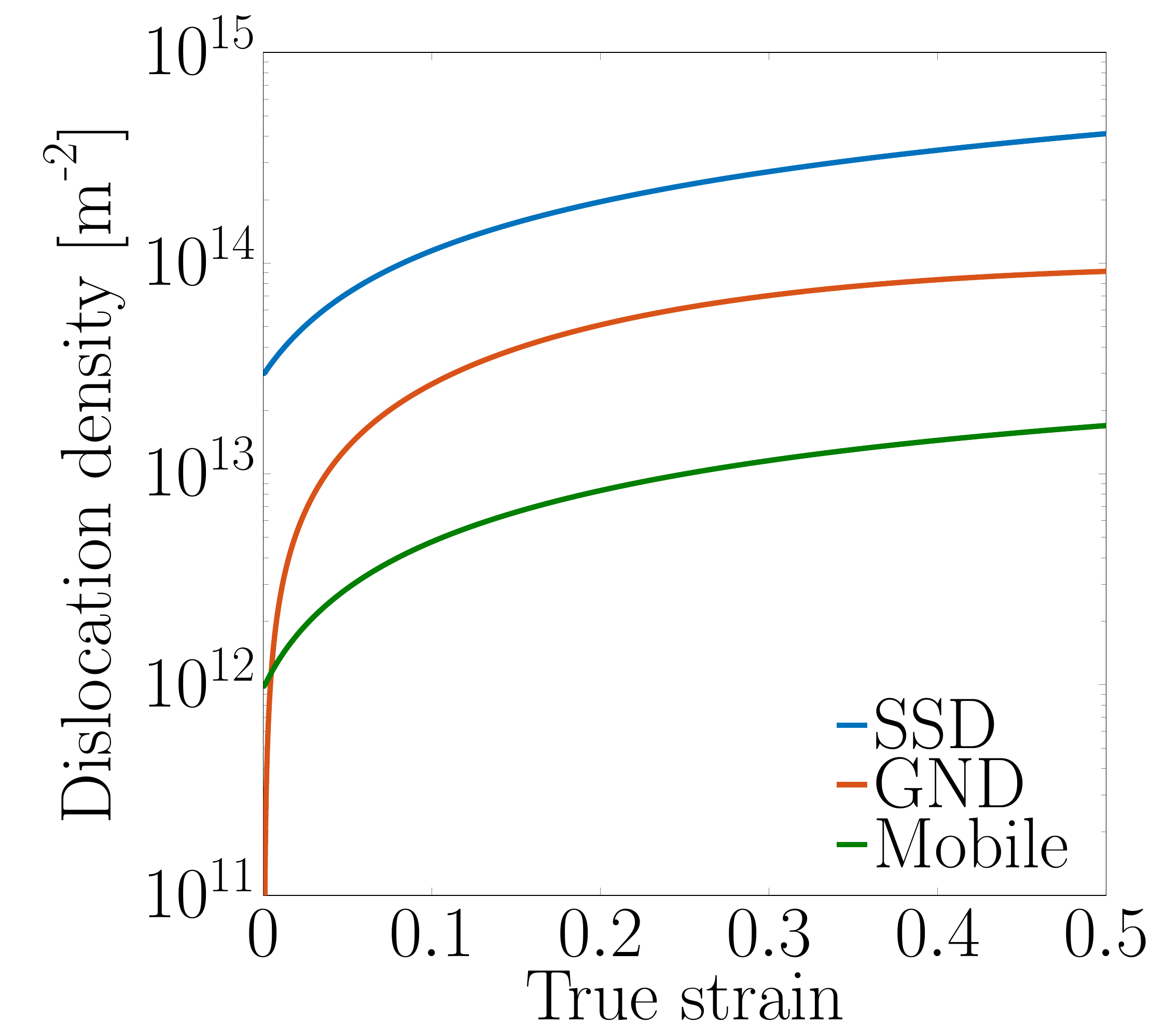}}
\hspace{14pt}
\subfigure[]{\label{fig:DD-723K}\includegraphics[width=0.38\textwidth]{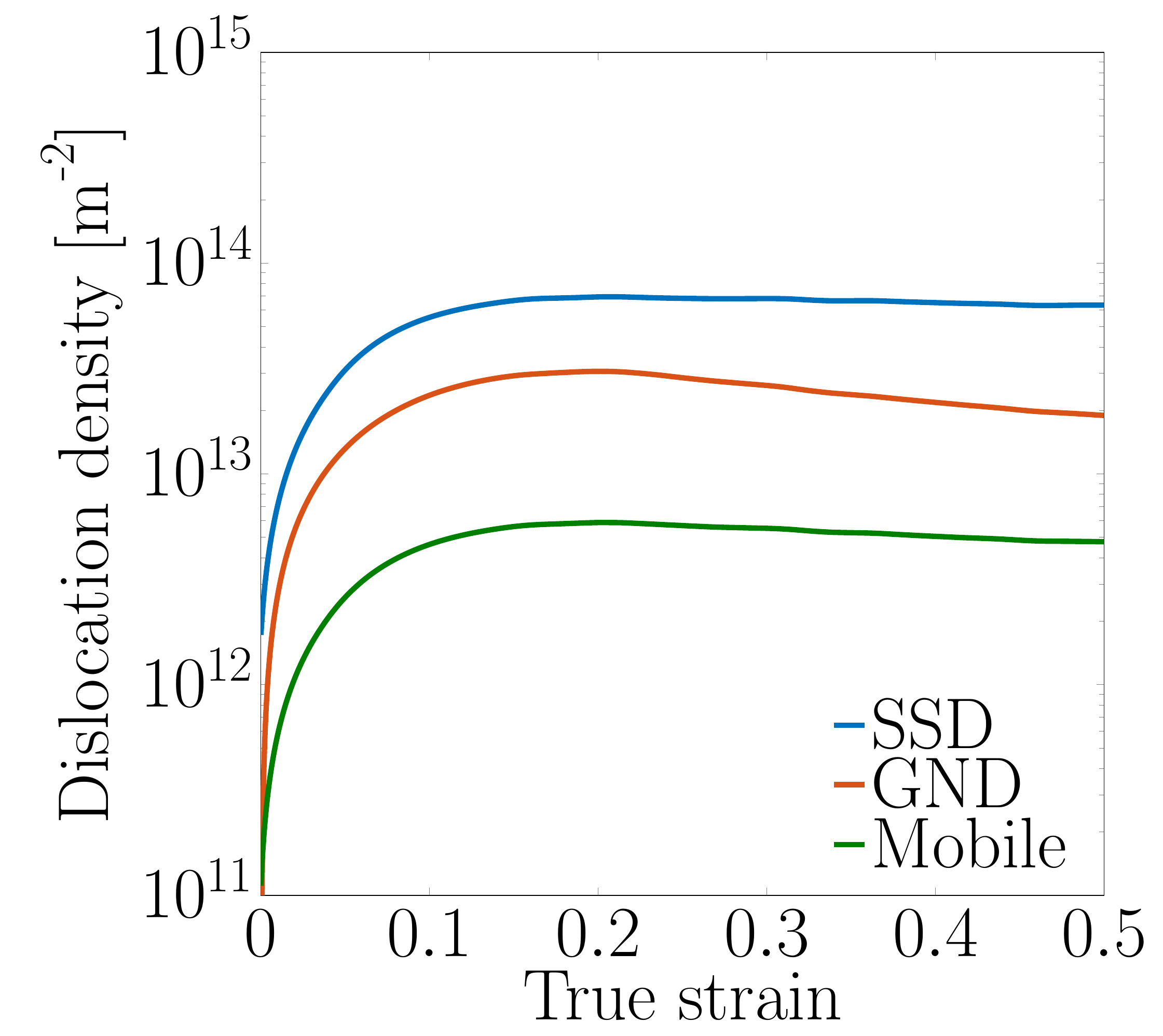}}
\caption{Comparison between experimental \citep{wusatowska2002nucleation} and simulated (a) compression stress-strain curves
at 473 K and 723 K and (b) the hardening rate at 723 K, with the corresponding evolution of dislocation densities at (c) 473 K based on FFT-CP simulation and (d) 723 K based on the integrated (FFT-EVP + PF) simulation.}
\label{fig:Fittings}
\end{center}
\end{figure*}

Simulation was carried out after the experimental study of \cite{wusatowska2002nucleation}. The material was 99.99\% pure Cu (4N) with equiaxed grains exhibiting an average size of $\sim230$ $\mu$m. The texture was nearly
random with weak ($<2$ times) random $\langle 1\:1\:1 \rangle$ and $\langle1\:0\:0\rangle$ components. The representative volume element (RVE) in our simulation was produced by DREAM3D \citep{groeber2014dream}.
\begin{figure*}[!ht]
\begin{center}
\subfigure[]{\label{fig:gID-initial}\includegraphics[width=0.17\textwidth]{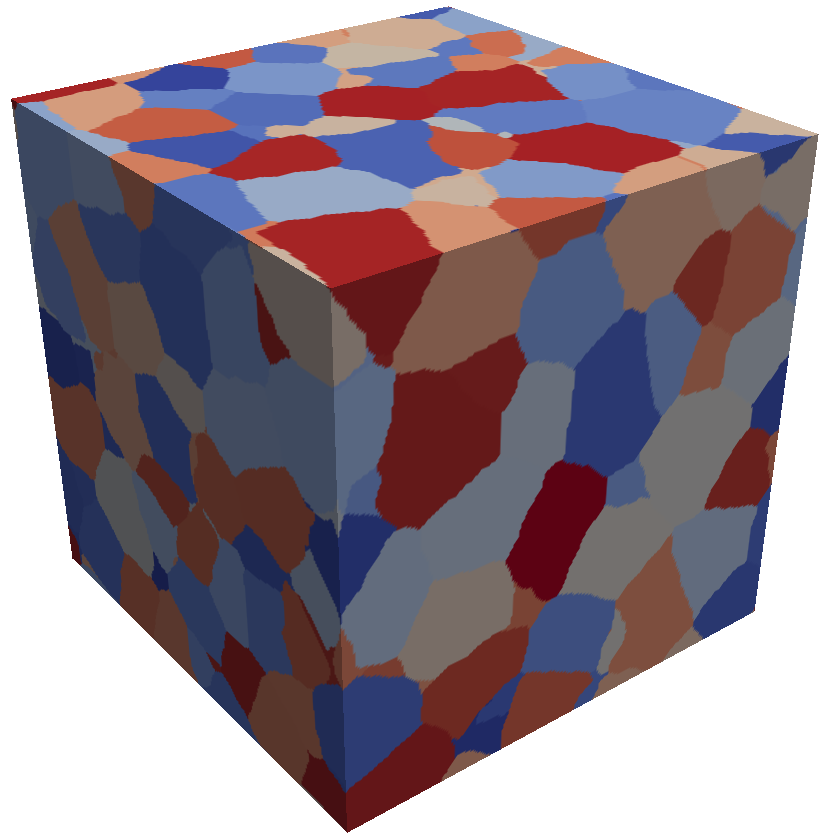}}
\subfigure[]{\label{fig:IPF-initial}\includegraphics[width=0.23\textwidth]{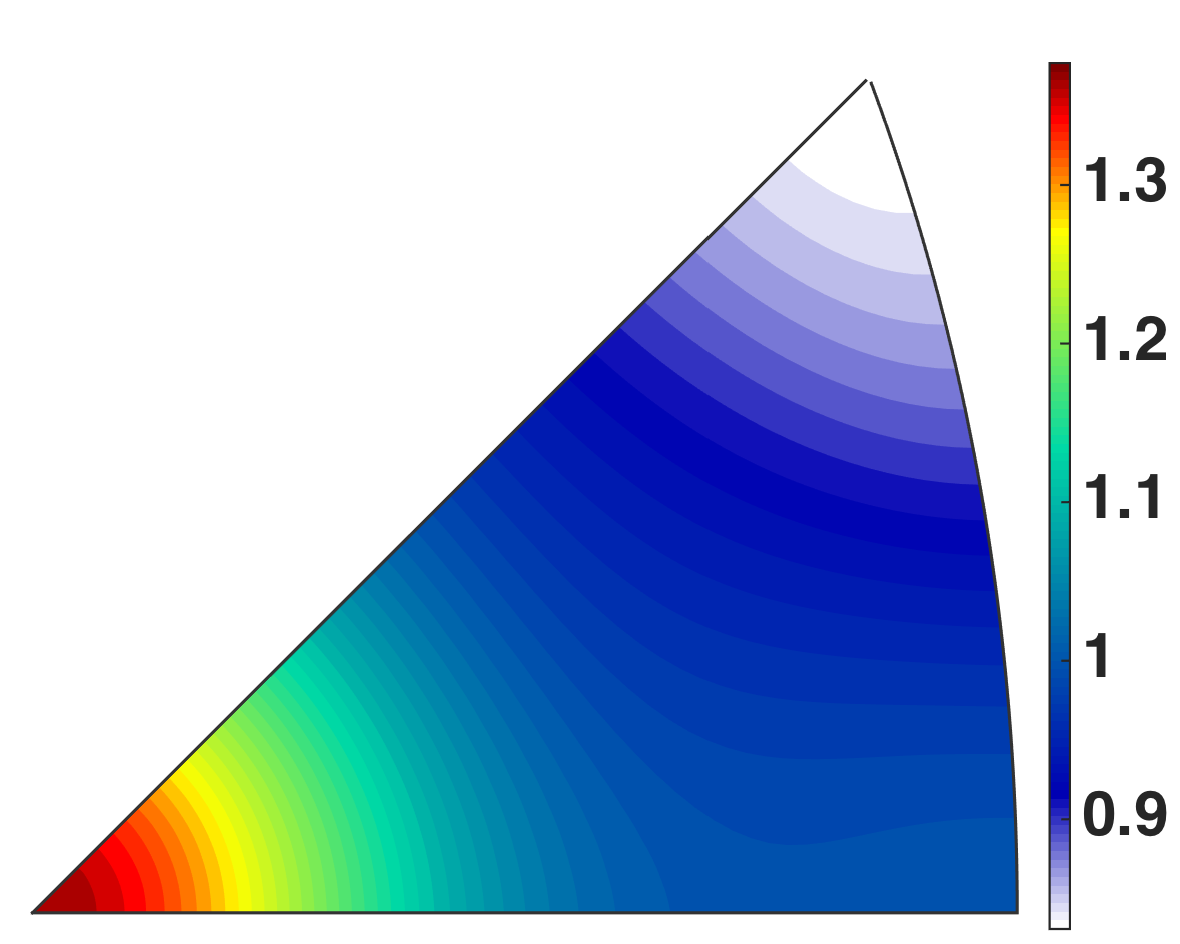}}
\subfigure[]{\label{fig:IPF-middle}\includegraphics[width=0.23\textwidth]{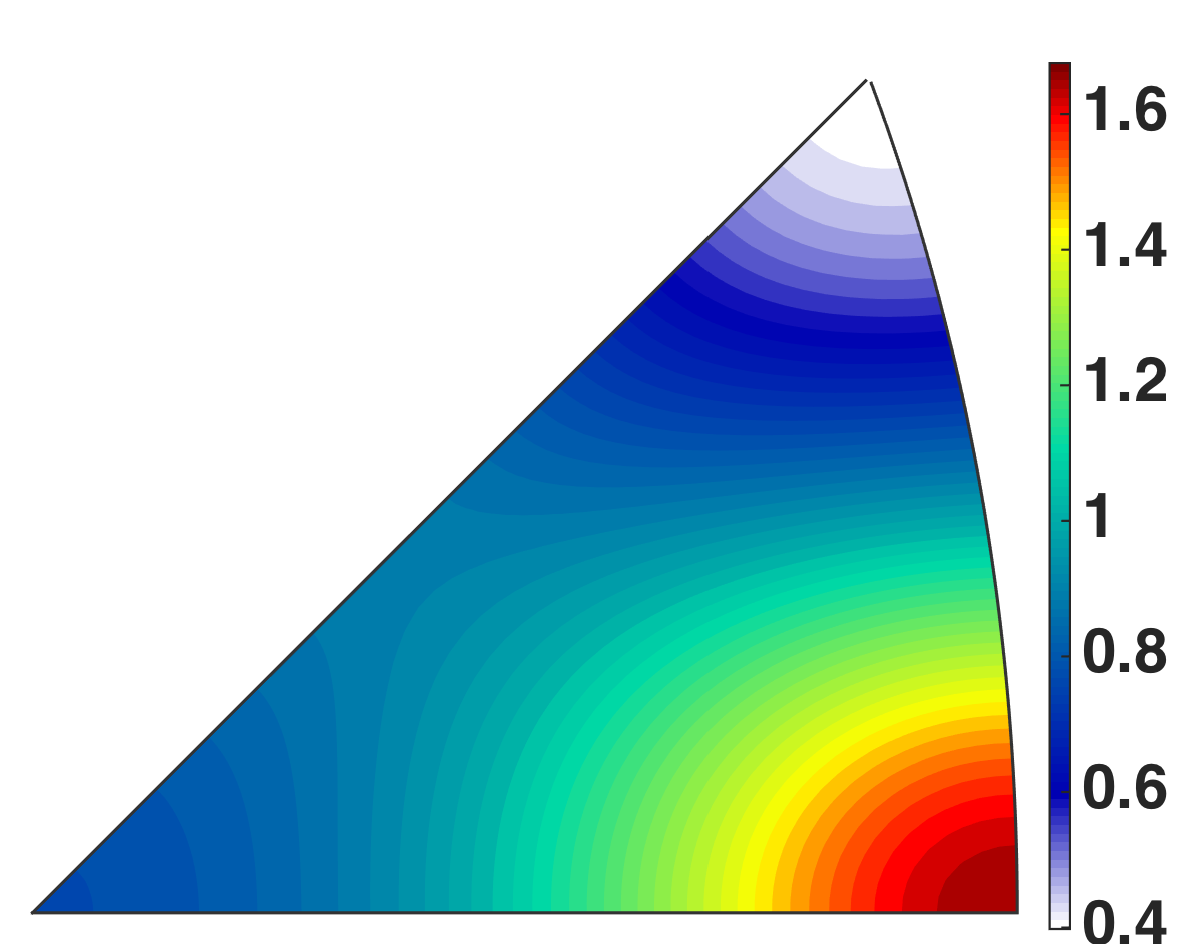}}
\subfigure[]{\label{fig:IPF-final}\includegraphics[width=0.23\textwidth]{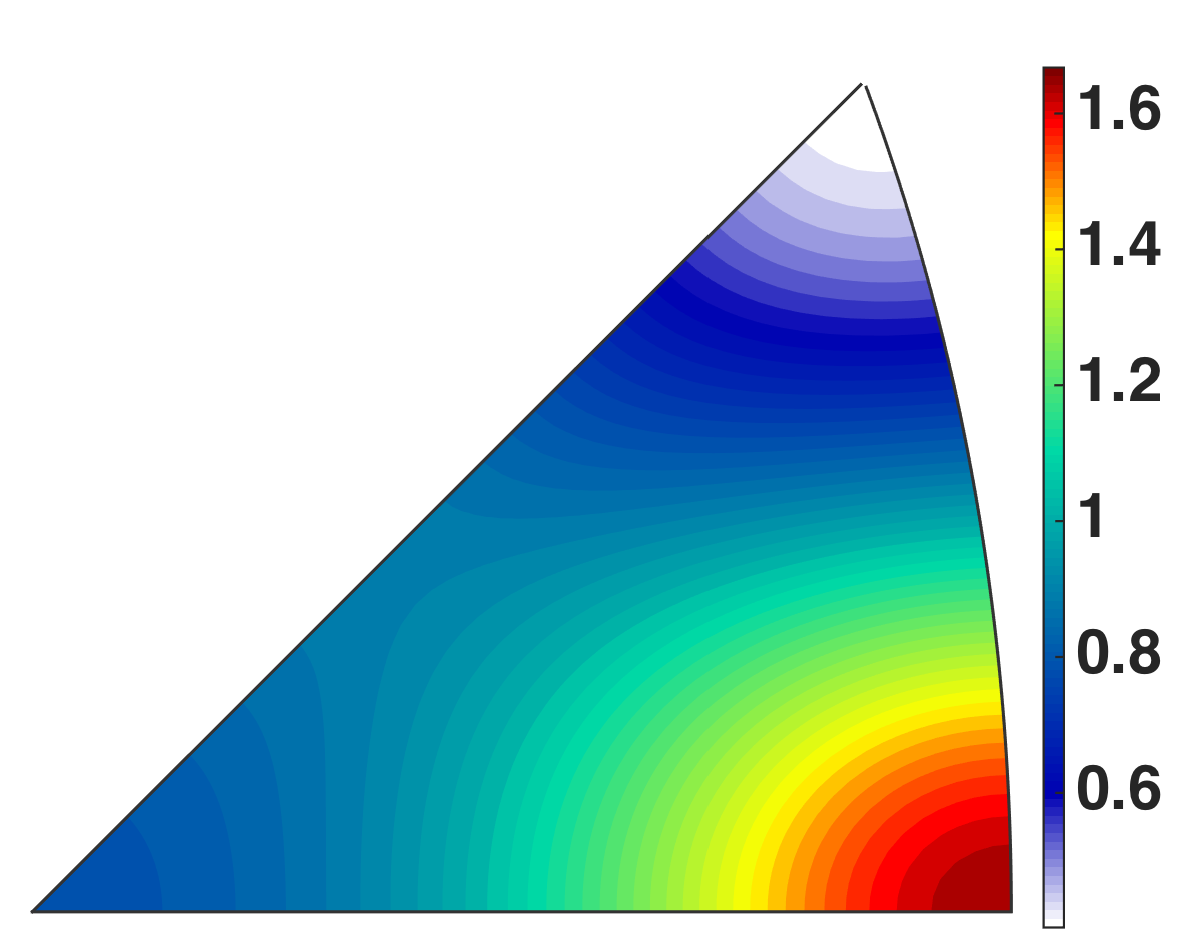}}
\caption{The input grain structure is randomly colored in (a) with the corresponding inverse pole figure (IPF) shown in (b)
(the preferred orientation is parallel to the compression axis).
IPFs at $50\%$ strain stages are shown for (c) a standalone FFT-EVP and (d) an integrated (FFT-EVP + PF)
simulations at 723 K,
of which the stress-strain curves are shown in Fig. \ref{fig:SSs}.}
\label{fig:IPFs}
\end{center}
\end{figure*}
The RVE contained 191 grains and the grid size $l_0=21.4$ $\mu$m was chosen such that the initial average grain size matches the experimental value. The initial grain structure and inverse pole figure (IPF) are shown in
Figs. \ref{fig:gID-initial} and \ref{fig:IPF-initial}, respectively.
\begin{table*}\centering
\caption{Parameters of dislocation-based constitutive model for copper.
($\dagger$\footnotesize{\cite{frost1982deformation}};
$\ddagger$\footnotesize{\cite{davis2001copper}}.)}
\label{tab:parameters-disl}
\vspace{+10pt}
\ra{1.3}
\begin{tabular}{@{}lclcl@{}}\toprule
Symbol && Value && Meaning\\ \midrule
$Q_{\rm slip}$ && $3.3\times10^{-19}$ J $\dagger$ && Activation energy barrier for slip\\
$Q_{\rm bulk}$ && $3.51\times10^{-19}$ J $\ddagger$ && Activation energy barrier for climb\\
$l_0$ && 21.4 $\mu$m && Grid size for a $64^3$ RVE containing 191 grains\\
$\xi$ && 200 && Constant for effective length of GND\\
$c_1$ && 0.5 && Constant for passing stress\\
$c_2$ && 3.0 && Constant for jump width\\
$c_3$ && 3.0 && Constant for obstacle width\\
$c_4$ && $8.0\times10^7$ ${\rm m}^{-1}$ && Constant for lock forming rate\\
$c_5$ && 10.0 && Constant for athermal annihilation rate\\
$c_6$ && $3.0\times10^{10}$ ${\rm m}^{-1}$ && Constant for thermal annihilation rate\\
$c_7$ && $7.0\times10^{-28}$ ${\rm m}^5{\rm s}^{c_8-1}$ && Constant for dipole forming rate\\
$c_8$ && 0.24 && Constant for nonlinear climb\\
\bottomrule
\end{tabular}
\end{table*}
Experimentally DRX was not observed in polycrystalline 4N Cu at 473K at strains $<50\%$ \citep{wusatowska2002nucleation}. Therefore a standalone FFT-EVP without PF coupling was used to calibrate
 the dislocation model. The RVE was subjected to simple compression at a strain rate of $1.6 \times 10^{-3}$/s to match the
 experimental conditions. The temperature dependence of the model was verified against the early part of the stress strain curve for 4N Cu at 723 K before the onset of dynamic recrystallization (See Fig.  \ref{fig:SSs}). The resulting
 parameters are listed in Table \ref{tab:parameters-disl} and the corresponding simulated stress-strain
curve is shown in Fig. \ref{fig:SSs}. It needs to be pointed out that
the initial SSD density is taken as $2.5\times10^{12}$ $\rm{ m}^{-2}$ at 473 K and
$2.8\times10^{11}$ $\rm{ m}^{-2}$ at 723 K. This is found necessary in order to account for
the significant difference between the initial yield points of the experimental stress-strain curves shown in Fig. \ref{fig:SSs}.
One possible reason is that effects such as grain boundary strengthening and/or
temperature-dependent activation volume have been ignored in the current study. The evolution of simulated dislocation densities are shown in Fig. \ref{fig:DD-473K},
which as expected shows that the mobile dislocation population is only a small fraction of that of the immobile dislocations.

With the calibrated constitutive law in hand, we now integrate FFT-EVP with
our nucleation model and PF model to implement the complete DRX simulation as depicted in Fig. \ref{fig:Model-CMS}.
Experimental values were used for GB energy and mobility \citep{murr1975interfacial, vandermeer1997grain} while the other nucleation and PF parameters were fit to the experimental stress/strain data (see Table \ref{tab:parameters-nucl}). \hl{The resulting fit reproduces both the experimental stress-strain curve and hardening rate (HR) evolution at 723 K with good agreement}, as shown,
respectively, in Fig. \ref{fig:SSs} and Fig. \ref{fig:HRs}. \hl{In the standalone crystal plasticity simulation, the HR
monotonically decreases to zero due to the dynamic recovery (i.e., annihilation terms
in Eq. \mbox{(\ref{eq:SSDEvolution})}). In the integrated simulation, the additional softening due to DRX
starts to produce a slight deviation on the HR curve around $10\%$ strain and gradually macroscopic softening, i.e.,
a negative slope on the stress-strain curve (or equivalently negative HR).
The corresponding evolution of average dislocation densities are shown in Fig. \mbox{\ref{fig:DD-723K}}, where
the average density of SSD exhibits a significant decrease
during DRX and is a major contributor to the softening.} As will be shown later, the softening
is mainly attributed to the local stress relaxation and redistribution in the neighborhood of newly nucleated and growing grains.
It should be noted that all model parameters were exclusively fit to the macroscopic stress-strain curves,
and the resulting evolution of dislocation densities is a key prediction of the coupled simulations. 

\begin{table*}\centering
\caption{Material properties and parameters of nucleation and phase-field models for copper.
($\dagger$\footnotesize{\cite{murr1975interfacial}};
$\ddagger$\footnotesize{\cite{vandermeer1997grain}}.)}
\label{tab:parameters-nucl}
\vspace{+10pt}
\ra{1.3}
\begin{tabular}{@{}lclcl@{}}\toprule
Symbol && Value && Meaning\\ \midrule
$\omega_{\rm gb}$ && $0.625$ ${\rm J}/{\rm m}^2$ $\dagger$ && Grain boundary energy\\
$M_{\rm gb}$ && $145$ ${\rm m}^4/({\rm MJ}\cdot{\rm s})$ $\ddagger$ && Grain boundary mobility\\
$\zeta$ && 0.25 && Constant for stored strain energy\\
$k_c$ && 300 $[10^{12}$ $\rm{ m}^{-2}]$ && Characteristic dislocation density for nucleation\\
$q$ && 4.4 && Exponent for Weibull distribution of nucleation rate\\
$s_{\rm nucl}$ && 0.05 && Accounting for the change of $k_c$ \\
$s_{\rm soften}$ && 0.9 && Fraction of GND inherited in DRX grains\\
\bottomrule
\end{tabular}
\end{table*}

\begin{figure*}[!ht]
\begin{center}
\subfigure[]{\label{fig:NumGrain}\includegraphics[width=0.32\textwidth]{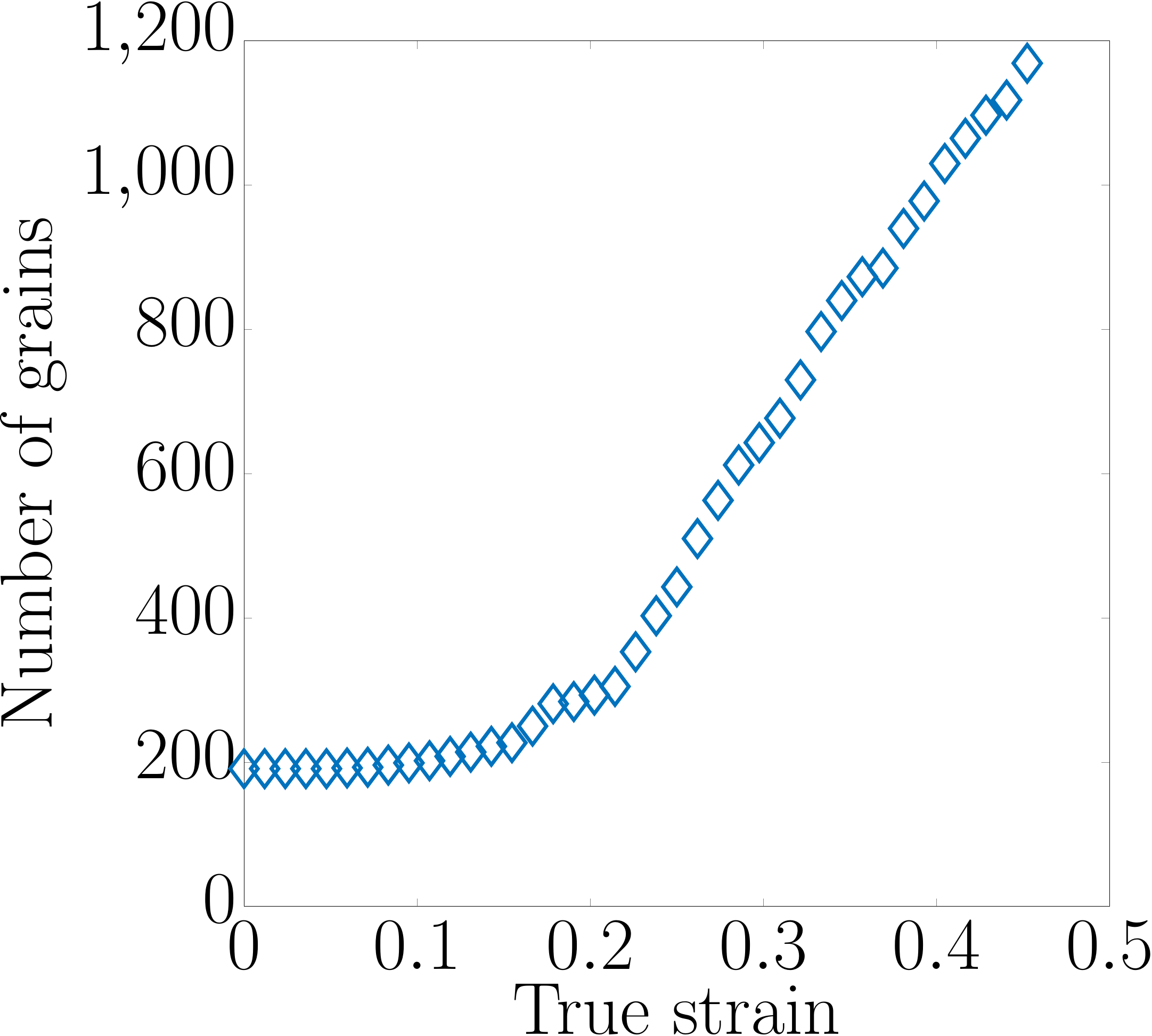}}
\subfigure[]{\label{fig:VolFrac}\includegraphics[width=0.32\textwidth]{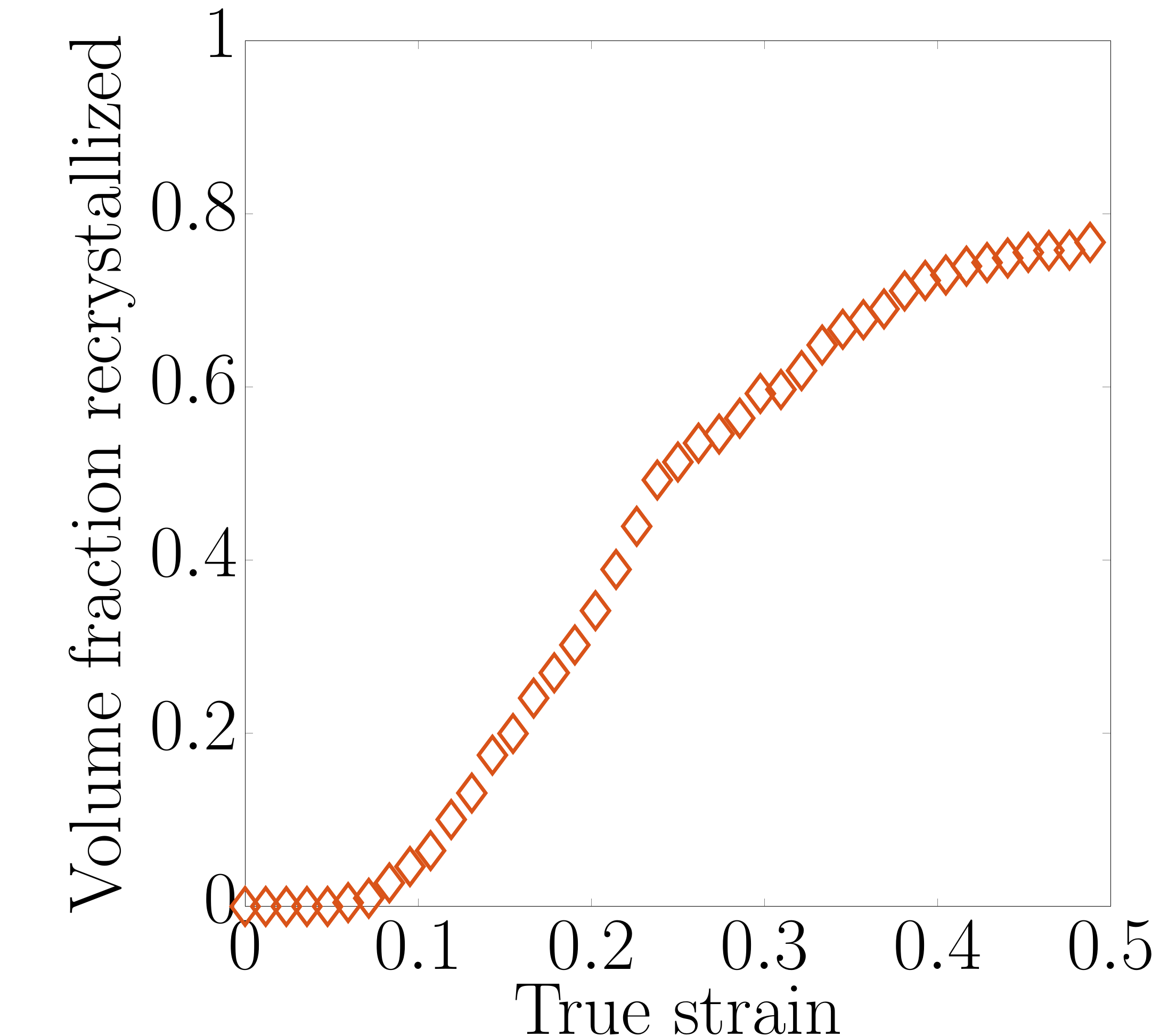}}
\subfigure[]{\label{fig:GrainSizeComp}\includegraphics[width=0.32\textwidth]{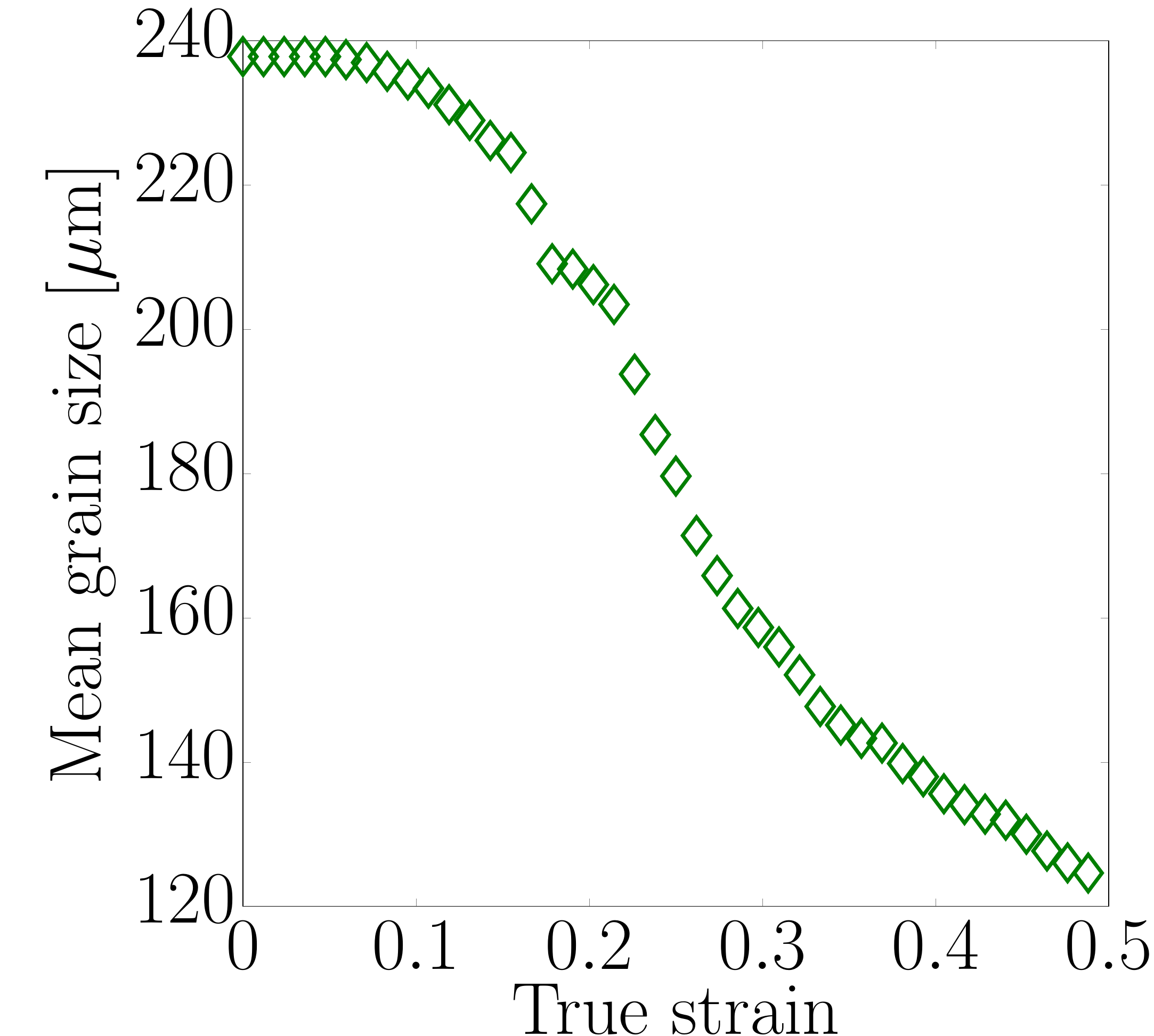}}
\caption{Evolution of (a) the total number of grains, (b) volume fraction of recrystallized grains and (c) mean grain size
during the integrated modeling at 723 K.}
\label{fig:Statistics}
\end{center}
\end{figure*}

\begin{figure*}[htb!]
\begin{center}
\subfigure[]{\label{fig:gID_03}\includegraphics[width=0.19\textwidth]{Figures/Refined_gID_S000000}}
\subfigure[]{\label{fig:gID_04}\includegraphics[width=0.19\textwidth]{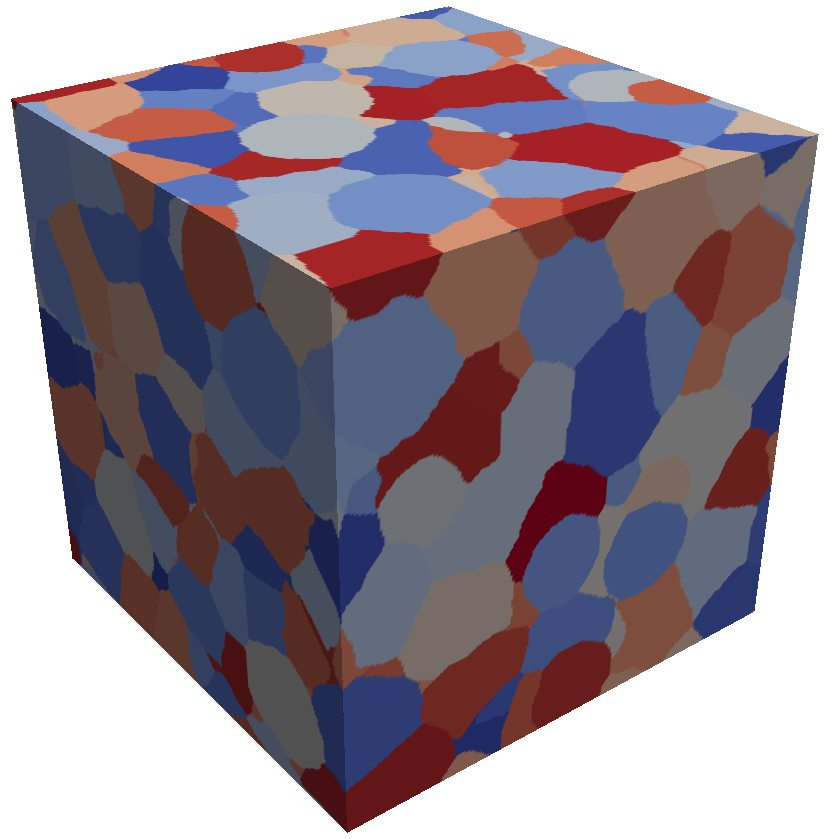}}
\subfigure[]{\label{fig:gID_05}\includegraphics[width=0.19\textwidth]{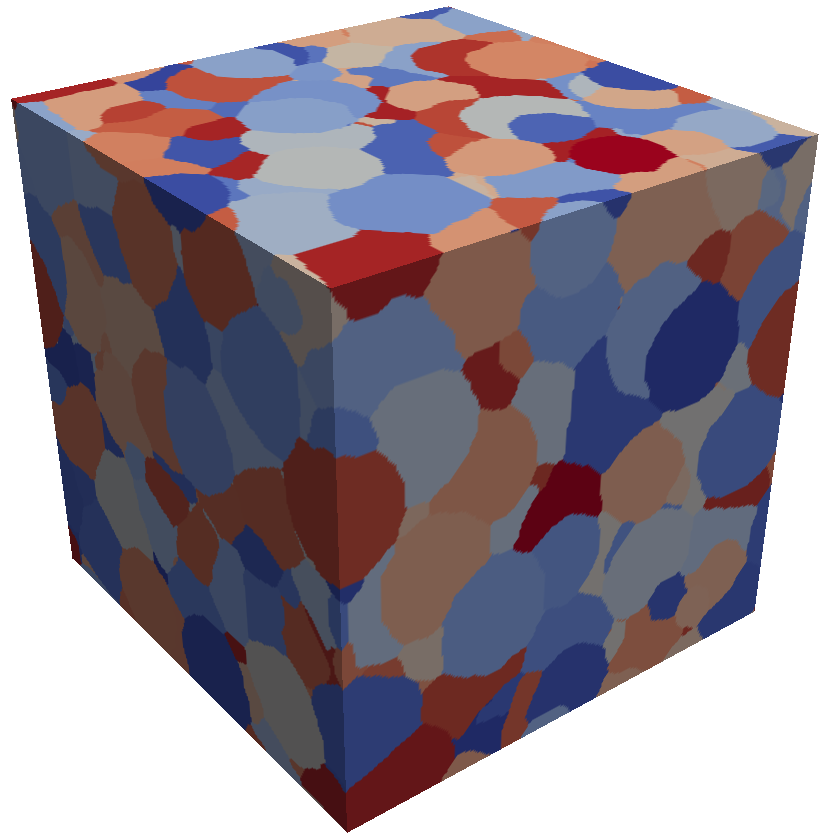}}
\subfigure[]{\label{fig:gID_06}\includegraphics[width=0.19\textwidth]{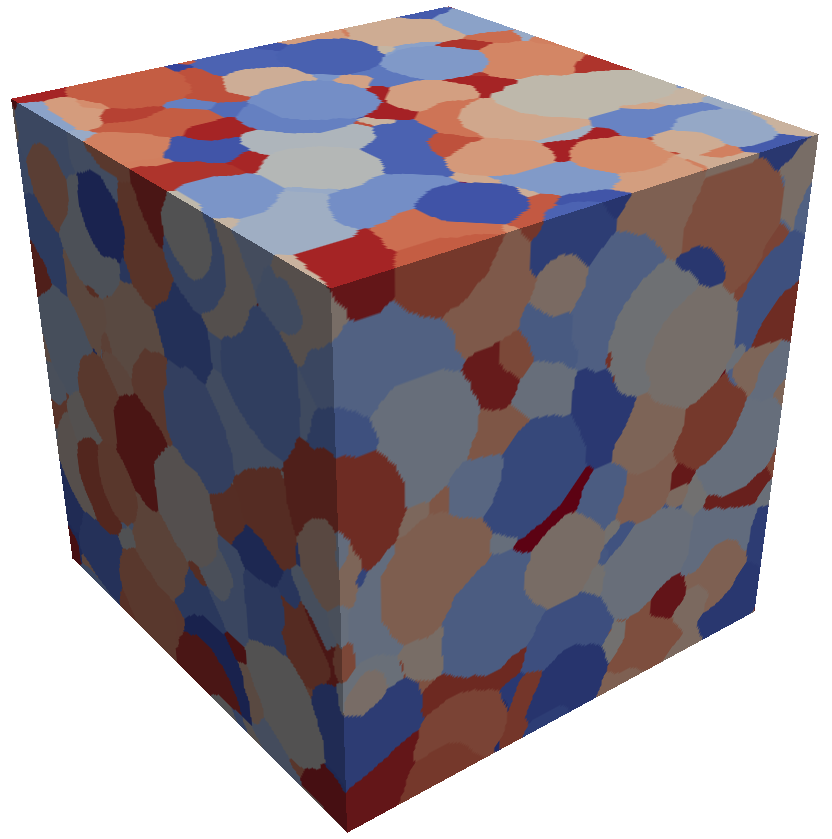}}
\subfigure[]{\label{fig:gID_07}\includegraphics[width=0.19\textwidth]{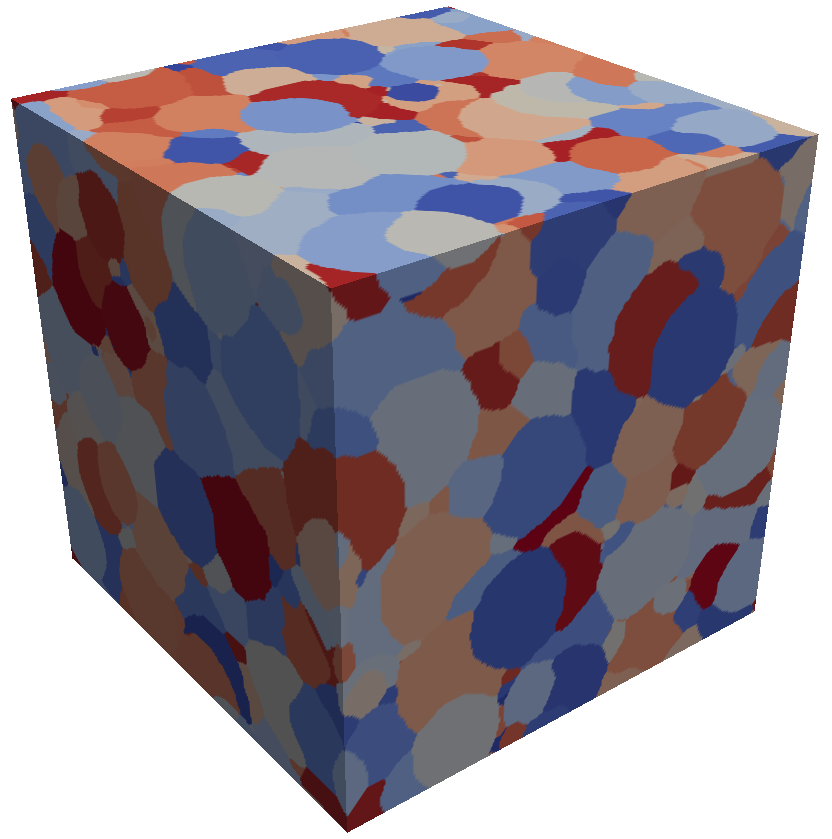}}\\
\caption{Grain map, \hl{colored by grain ID}, at the strain of (a) $0\%$, (b) $14.4\%$, (c) $26.3\%$, (d) $38.2\%$, and
(e) $50.0\%$ during the integrated modeling at 723 K. The DRX grains at different strains can be
identified by comparing the corresponding microstructures in (b)-(e) to that in (a).}
\label{fig:Evolution}
\end{center}
\end{figure*}

Another key prediction is the full-field microstructural information during DRX.
Figs. \ref{fig:Statistics} plots the evolution of total number of grains in the RVE, the volume fraction of recrystallized grains
and the mean grain size during the deformation. The overall
sigmoidal form of the recrystallized volume fraction agrees with the experiments \citep{blaz1983effect,rollett2004recrystallization}, and the continuous increase of the number of gains suggests grain refinement,
which agrees with the experiment as well.
The underlying grain structure evolution is shown in Fig. \ref{fig:Evolution} by plotting snapshots of grain maps during
the compression. It can be seen that
DRX grains have been formed at both grain boundaries and junctions. Note that the grain structure has changed
so significantly that the detailed DRX nucleation and growth
processes can only be achieved by more quantitative analysis of the simulation result,
which will be presented in the following Discussion section.
The simulation also shows that
new grains are mostly equiaxed, which is consistent with the corresponding experiment \citep{wusatowska2002nucleation}.
IPFs at the macroscopic strain of $50\%$ for the standalone FFT-EVP and integrated simulations at 723 K (Fig. \ref{fig:SSs})
are plotted in Figs. \ref{fig:IPF-middle} and \ref{fig:IPF-final}, respectively. The exhibited
formation of $\langle 1\:0\:1\rangle$ fiber texture is consistent
with the experiment \citep{wusatowska2002nucleation}. In addition, DRX essentially places no difference in terms of the texture,
which is also consistent with the experimental conclusion that DRX only causes slightly weakening of the
$\langle 1\:0\:1\rangle$ fiber texture \citep{wusatowska2002nucleation}.

\section{Discussion}
\label{sec:disc}

\subsection{Critical strain of DRX}
\label{sec:critical_strain}
The initiation of DRX is well documented to occur before the strain $\varepsilon_p$ that
corresponds to the peak stress \citep{sakai1984overview}.
This threshold strain $\varepsilon_c$, known as the critical strain of DRX, is a very fundamental DRX parameter
for practical purpose. However, it can only be determined
under isothermal, constant strain rate conditions, involving a large number of interrupted tests
\citep{manonukul1999initiation}. It is expected that experimental identification of $\varepsilon_c$
based on 2D cross-section should be always larger than the actual strain
corresponding to the very first DRX event, for sufficient volume fraction of DRX grains is required
to yield statistically observable new grains in a randomly chosen 2D cross-section.
Our integrated modeling provides an accurate way of predicting $\varepsilon_c$
by simply examining the first DRX nucleation event.
Table \ref{tab:CriticalStrainDRX} lists our simulation result together with experimental results performed on
polycrystalline 3N Cu at a slightly different temperature (700 K) with various strain rates.
\begin{table*}\centering
\caption{The critical strain to initiate DRX in polycrystalline copper. Experimental data are taken from
\citep{manonukul1999initiation}.
(\textsuperscript{\textdagger}\footnotesize{The simulation prediction should always be smaller than the experimental measurement.})}
\label{tab:CriticalStrainDRX}
\vspace{+10pt}
\ra{1.3}
\begin{tabular}{@{}lcccccc@{}}\toprule
&\phantom{abc} & \multicolumn{3}{c}{Experiments, 700 K} &
\phantom{abc}& \multicolumn{1}{c}{Simulation, 723 K}\\
\cmidrule(lr){3-5} \cmidrule(lr){7-7}
Strain rate (${\rm s}^{-1}$)&& $5\times10^{-4}$ & $5\times10^{-3}$ & $5\times10^{-2}$ && $1.6\times10^{-3}$
\\ \midrule
$\varepsilon_c$&&0.15&0.19&0.23&&0.06\textsuperscript{\textdagger}\\
$\varepsilon_p$&&0.22&0.30&0.36&&0.22\\
$\varepsilon_c/\varepsilon_p$&&0.68&0.63&0.64&&0.27\textsuperscript{\textdagger}\\
\bottomrule
\end{tabular}
\end{table*}
As expected our predicted $\varepsilon_c$ is significantly smaller than the experimental measurement, which
should only serve as
the upper bound of the actual $\varepsilon_c$. Our integrated modeling suggests that in reality the initiation of DRX
may occur at a much earlier stage than what is revealed by the existing experimental technique and mean-field theory.
\begin{figure*}[htb!]
\begin{center}
\subfigure[]{\label{fig:DislDists}\includegraphics[width=0.42\textwidth]{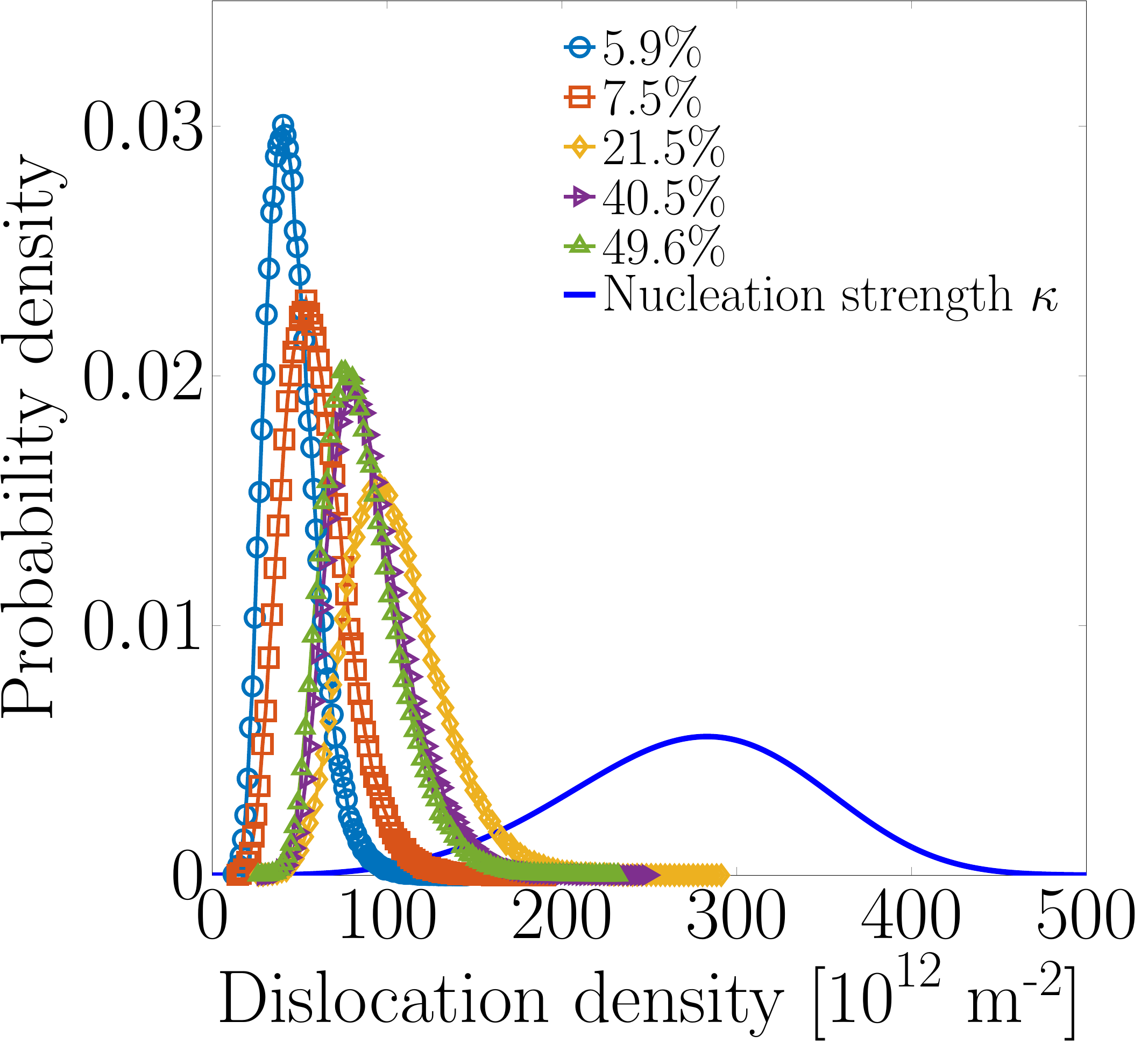}}
\subfigure[]{\label{fig:VolPer_DRXNucl}\includegraphics[width=0.43\textwidth]{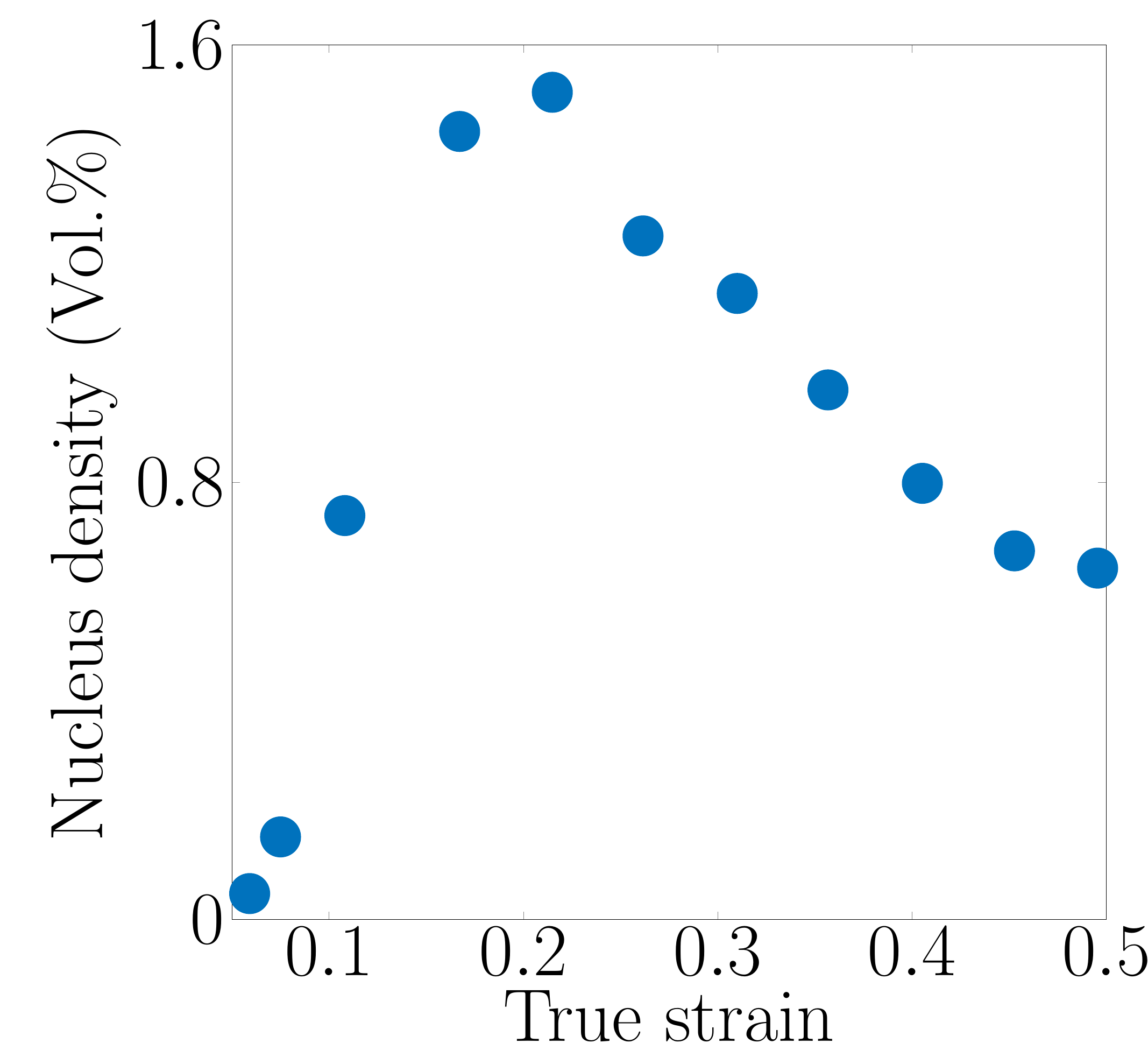}}
\caption{(a) The dislocation density distributions at various strain levels prior DRX events and (d) the
corresponding expected nucleus density (i.e., volume percentage of DRX nucleation at mesoscale voxel level). The nucleation strength
distribution (Fig. \ref{fig:NuclStrength-prob}) is also plotted in (a).}
\label{fig:StatNuclAnaly}
\end{center}
\end{figure*}
\hl{
In Fig. \mbox{\ref{fig:DislDists}}, the dislocation density distributions at various
strain levels prior to DRX nucleation are plotted. With increasing applied strain the dislocation density distribution shifts towards towards higher dislocation densities and broadens. After an applied strain of $\approx 21.5\%$ ($\sim \varepsilon_p$),
however, the mean of the distribution shifts back towards
lower values, indicating a macroscopic softening. In addition, the distribution of dislocation density
overlaps with the distribution of nucleation strength only at tails (Fig. \mbox{\ref{fig:DislDists}}), suggesting
that it is the material points with extreme values of dislocation density that actually controls the initiation of DRX.
Integrating the overlap between the two distributions, $\int f(k(x))\cdot f_\kappa(k)dk$, gives the expected nucleus density (i.e., volume percentage of mesoscale voxels undertaking DRX nucleation) at the given
deformation state as shown in Fig. \mbox{\ref{fig:VolPer_DRXNucl}}, which is a direct measurement of instantaneous nucleation rates.
It suggests that the DRX nucleation rate increases significantly once initiated and reaches
a maximum at $\sim\varepsilon_p$, followed by a gradual decrease. As discussed above, DRX is triggered by
extreme sites, which keep increase as the material continue to work-harden. Initially, the work-hardening
outweighs the softening due to recovery and DRX, and the nucleation rate reaches the maximum
at $\sim\varepsilon_p$, at which the work-hardening is balanced by the softening and the apparent hardening rate
becomes zero as shown in Fig. \mbox{\ref{fig:HRs}}. The post-$\varepsilon_p$ softening,
originated from the stress redistribution that will be discussed in Sec. \mbox{\ref{sec:softening}},
implies that the population of extreme sites will no longer increase (as the dislocation density distribution is shifting
towards lower values as shown in Fig. \mbox{\ref{fig:DislDists}}) and a macroscopic
stress-drop is developed as the dynamic system approaches the steady-state.
}

\subsection{Nucleation and growth of DRX grains}
\begin{figure*}[htb!]
\begin{center}
\subfigure[]{\label{fig:1stDRX-S1}\includegraphics[width=0.23\textwidth]{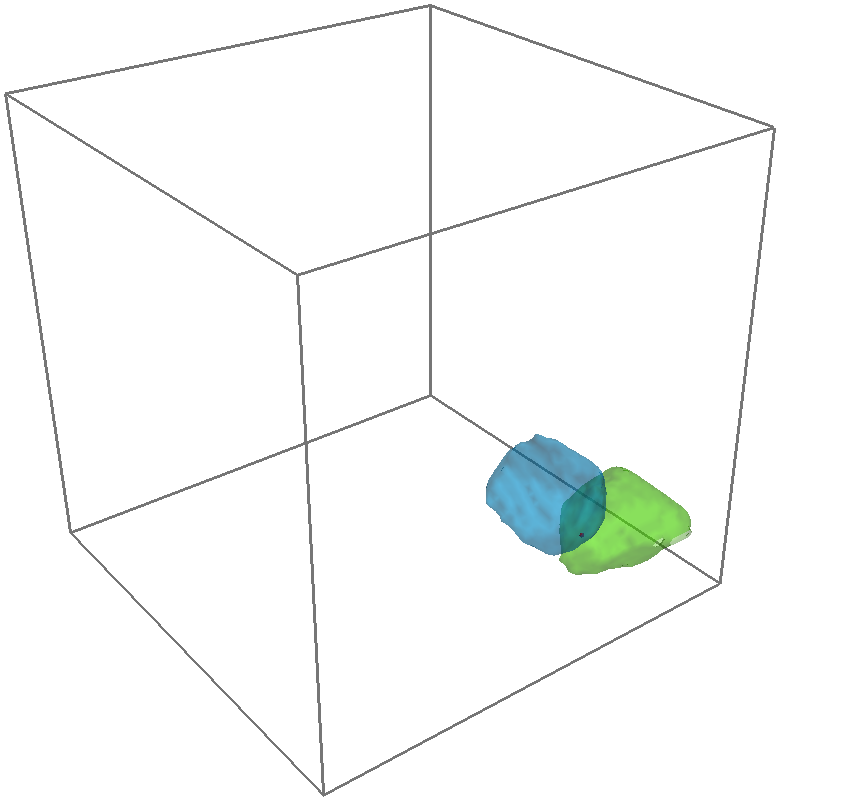}}
\subfigure[]{\label{fig:1stDRX-S2}\includegraphics[width=0.23\textwidth]{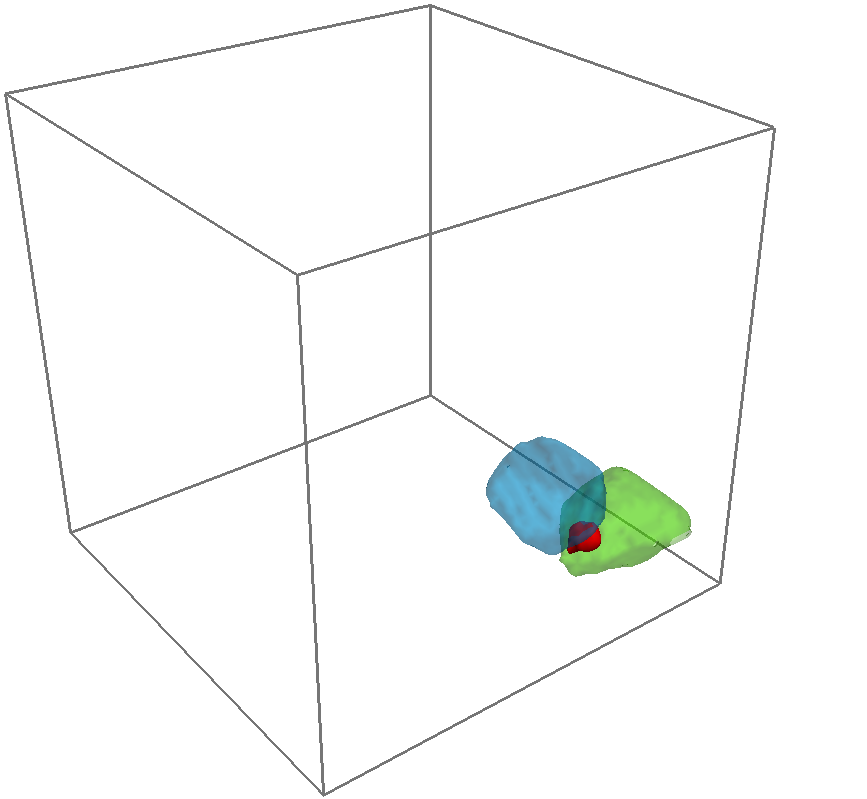}}
\subfigure[]{\label{fig:1stDRX-S3}\includegraphics[width=0.23\textwidth]{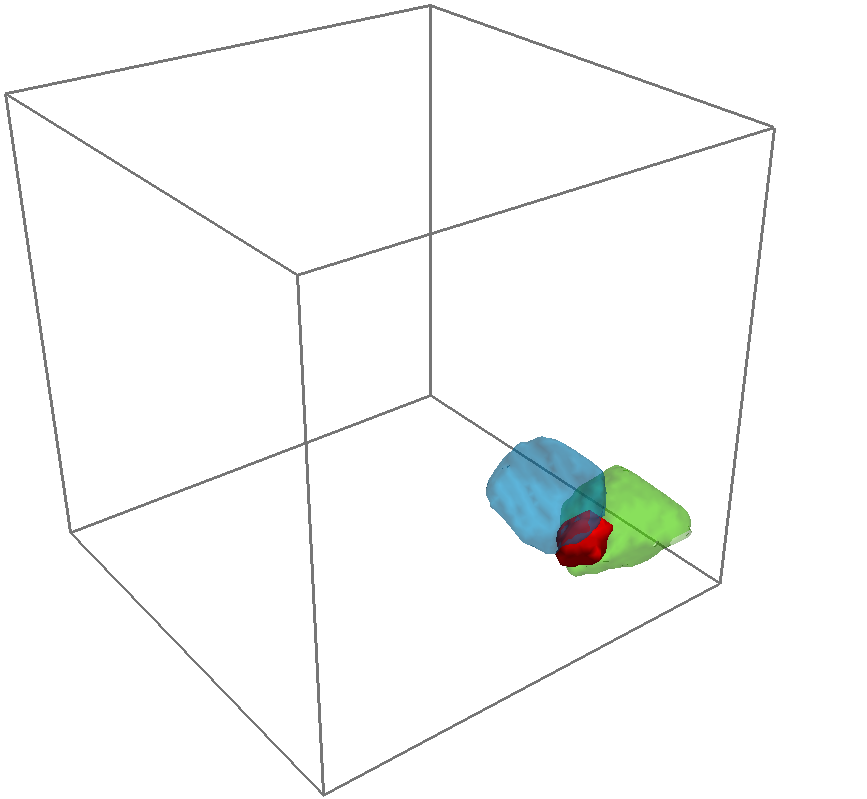}}
\subfigure[]{\label{fig:1stDRX-S4}\includegraphics[width=0.23\textwidth]{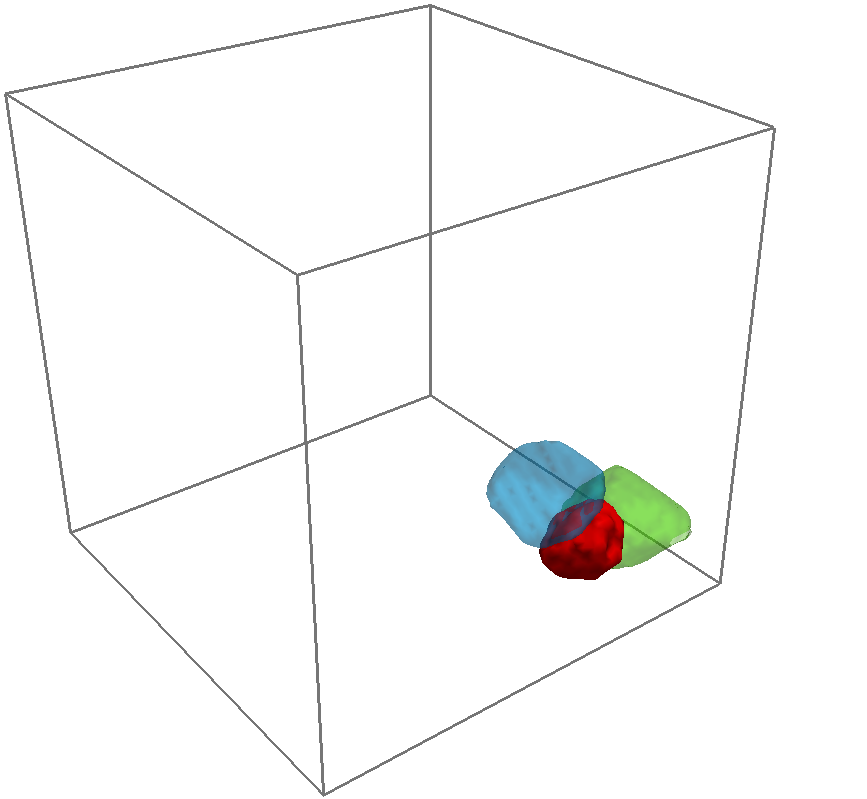}}
\caption{Growth of the first nucleated DRX (red grain) during the PF relaxation of the integrated modeling at PF time
$t=$ (a) 0 s, (b) 0.016 s,
(c) 0.021 s, and (d) 0.030 s. \hl{Here the initial stage (a) corresponds to the global (physical) time 37 s}
and the time step in FFT-EVP simulation is $\Delta t=0.03$ s.}
\label{fig:Growth-1stDRX}
\end{center}
\end{figure*}

The growth of the very first DRX grain is shown in Fig. \ref{fig:Growth-1stDRX}.
This DRX grain is formed at a normal grain boundary. In addition, the new
grain grows in a wedge-like fashion to maintain a triple line with the two old grains, wherein the equilibrium
triple junction configuration ($120^\circ$) does not need to be reached. For one thing,
grain growth is now mainly driven by the stored energy difference rather than the curvature; for another,
the dynamic nature of DRX is very likely to prevent the equilibrium configuration from being reached
before other nucleation and growth events come into play.

\begin{figure*}[htb!]
\begin{center}
\subfigure[]{\label{fig:2ndDRX-S1}\includegraphics[width=0.23\textwidth]{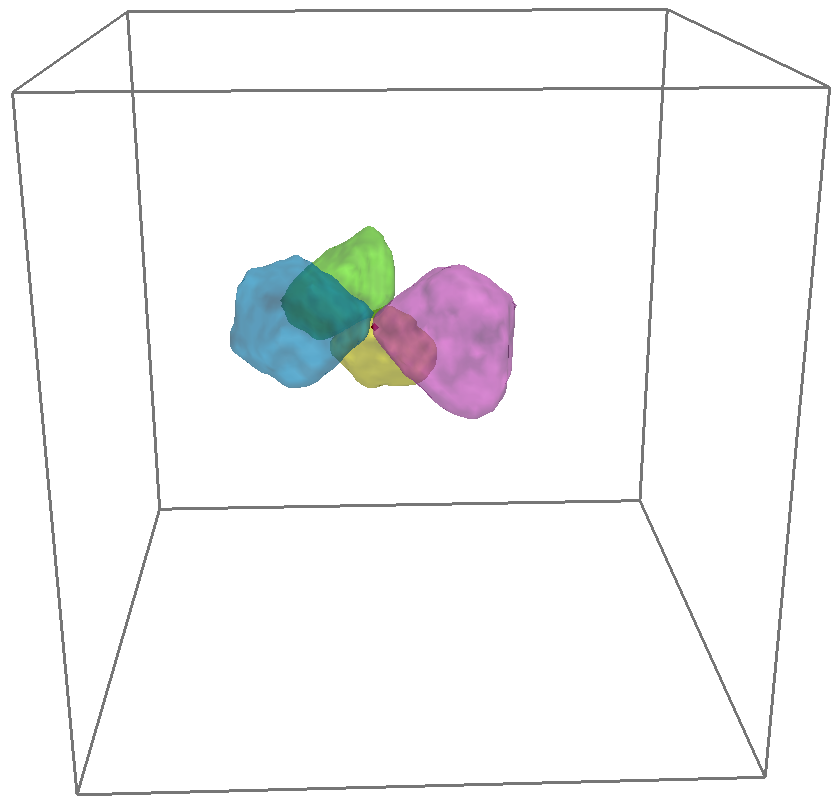}}
\subfigure[]{\label{fig:2ndDRX-S2}\includegraphics[width=0.23\textwidth]{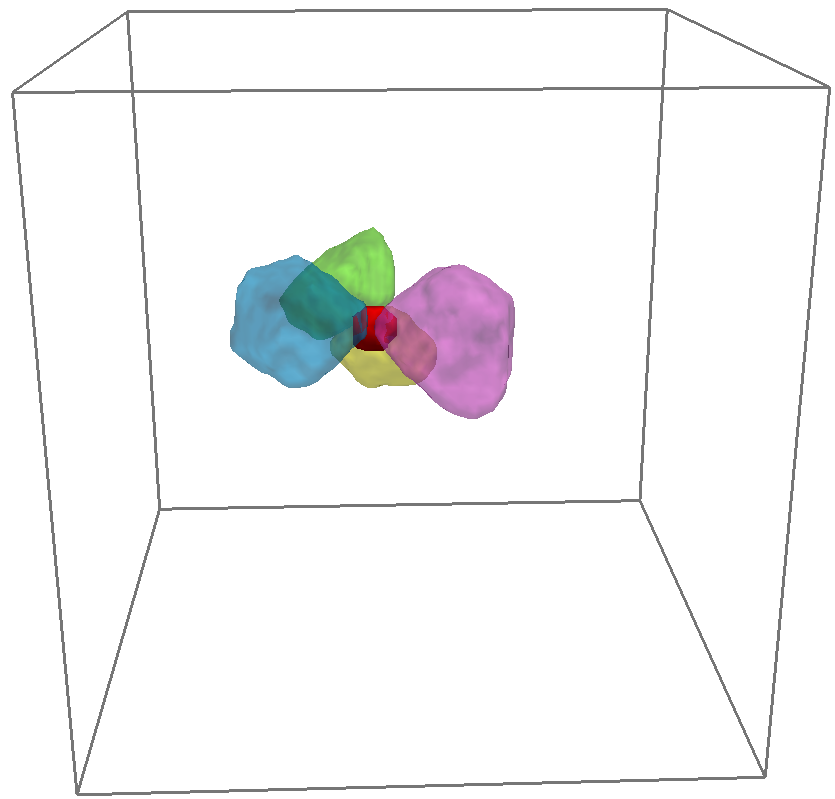}}
\subfigure[]{\label{fig:2ndDRX-S3}\includegraphics[width=0.23\textwidth]{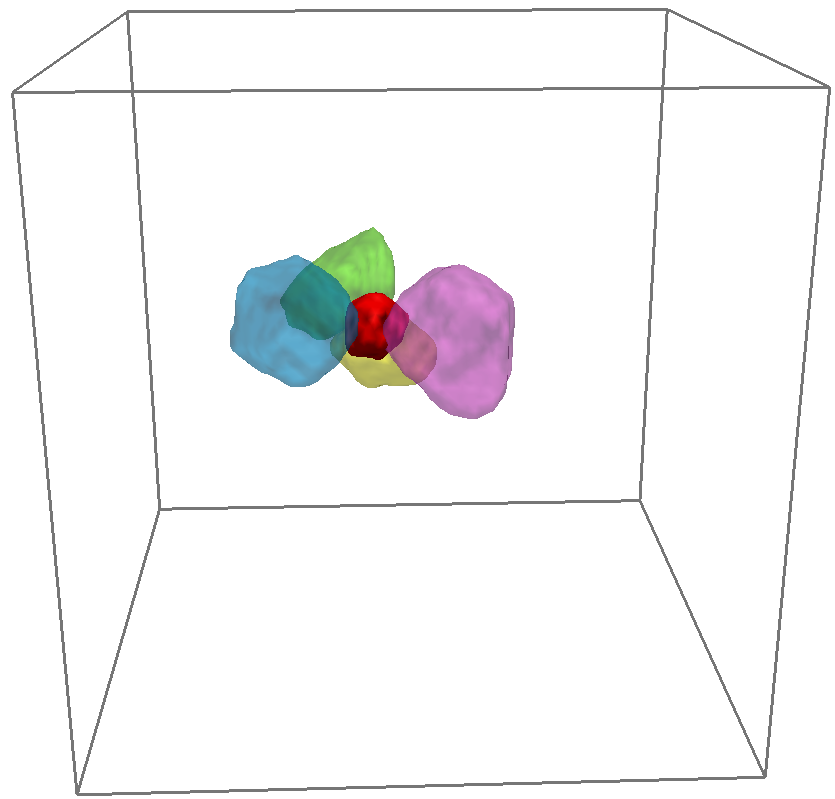}}
\subfigure[]{\label{fig:2ndDRX-S4}\includegraphics[width=0.23\textwidth]{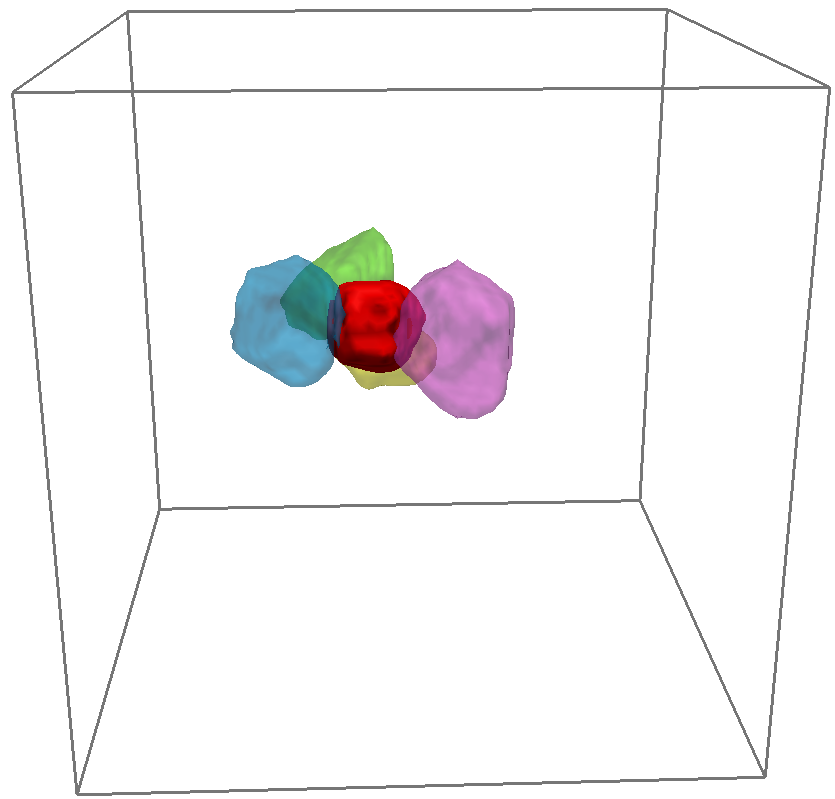}}
\caption{Growth of the a nucleated DRX (red grain) at a quadruple junction during the PF relaxation of
the integrated modeling at PF time
$t=$ (a) 0 s, (b) 0.016 s,
(c) 0.021 s, and (d) 0.030 s. \hl{Here the initial stage (a) corresponds to the global (physical) time 47 s} and
the time step in FFT-EVP simulation is $\Delta t=0.03$ s.}
\label{fig:Growth-2ndDRX}
\end{center}
\end{figure*}

\begin{figure*}[htb!]
\begin{center}
\subfigure[]{\label{fig:TripleLine-1}\includegraphics[width=0.23\textwidth]{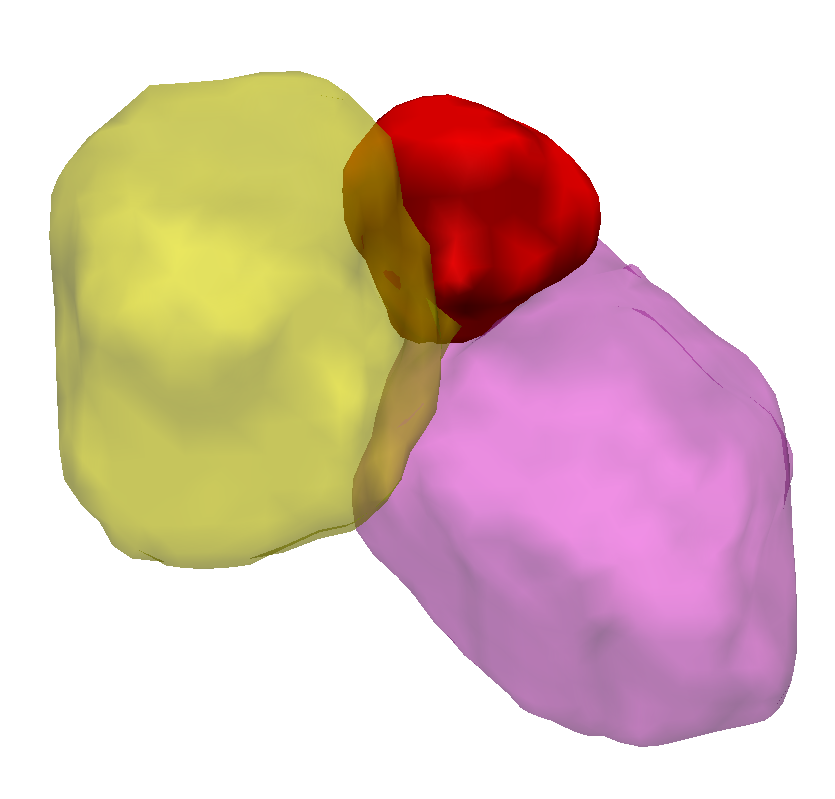}}
\subfigure[]{\label{fig:TripleLine-2}\includegraphics[width=0.23\textwidth]{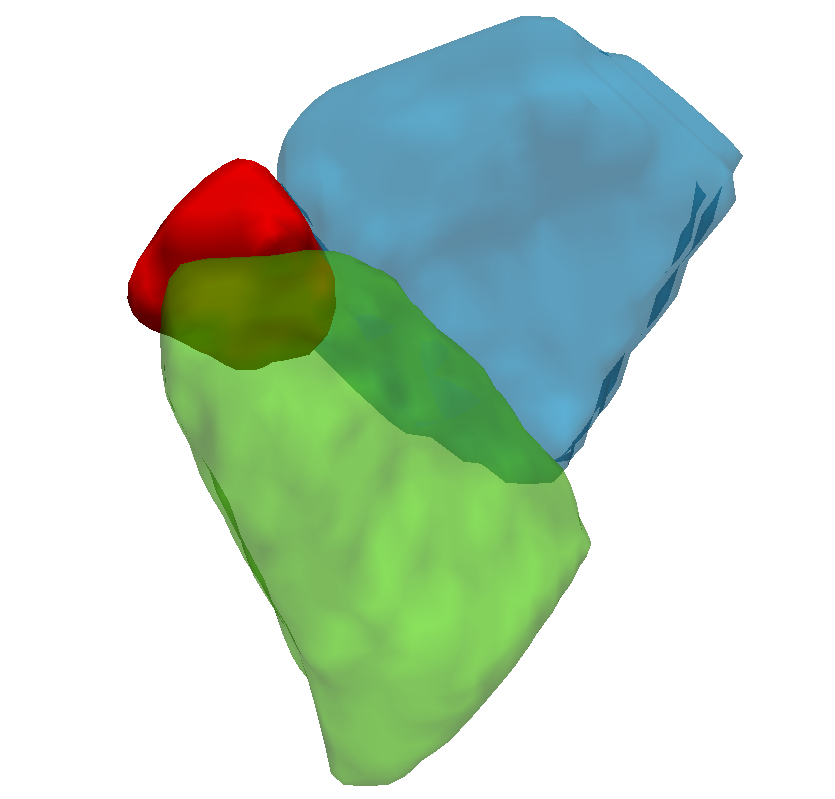}}
\subfigure[]{\label{fig:TripleLine-3}\includegraphics[width=0.23\textwidth]{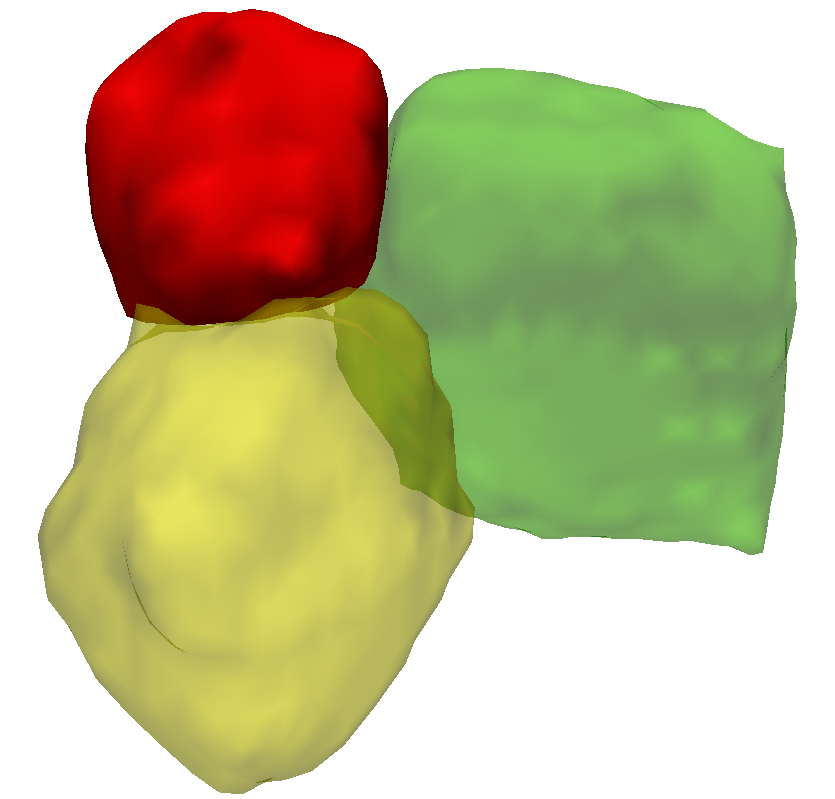}}
\subfigure[]{\label{fig:TripleLine-4}\includegraphics[width=0.23\textwidth]{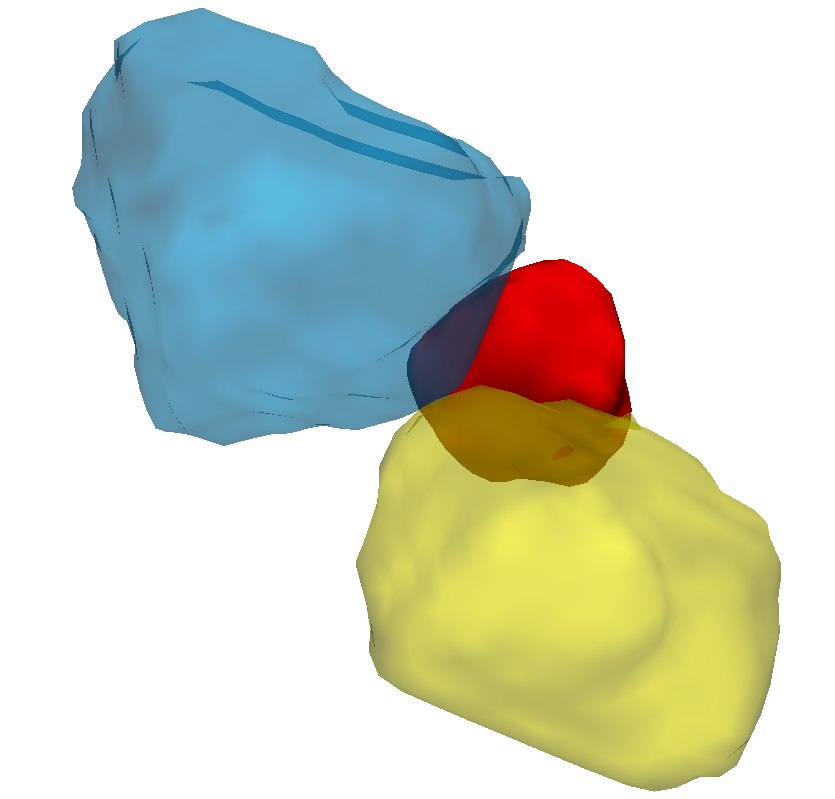}}
\caption{Individual triple junctions formed by every three contacting grains at a DRX grain formed
at quadruple junction as shown in Fig. \ref{fig:Growth-2ndDRX}. The viewpoint has been adjusted
accordingly for the best illustration.}
\label{fig:TripleLines}
\end{center}
\end{figure*}

Apart from nucleation at grain boundaries, experiments have indicated that triple junctions may be the preferential
nucleation sites for DRX grains \citep{miura2005nucleation}. However, as has also been pointed out
by \cite{miura2005nucleation}, from 3D point of view, quadruple junctions should be more preferred over triple junctions
in terms of DRX nucleation due to the fact that
stress or strain can concentrate more easily at quadruple junctions owing to its non-equilibrium
nature. In our simulation, we
have actually identified the nucleation of DRX at a quadruple junction,
as shown in Fig. \ref{fig:Growth-2ndDRX}.
A close inspection of the 3D structure as shown in Fig. \ref{fig:TripleLines}
suggests that as in the case of
nucleation and growth at a normal GB, the DRX grain formed at a quadruple junction also tends to maintain
a triple line with each pair of the neighboring grains by growing in a wedge-like fashion.

It would also be of great interest to analyze all nucleated DRX grains
in our integrated simulations and perform statistics on the nucleation at GB, triple junctions and quadruple junctions.
However, this is beyond the purpose of the current paper and we leave that as a topic of future work.

\subsection{Kinetics of DRX}

\begin{figure*}[htb!]
\begin{center}
\subfigure[]{\label{fig:AvramiExpo}\includegraphics[width=0.42\textwidth]{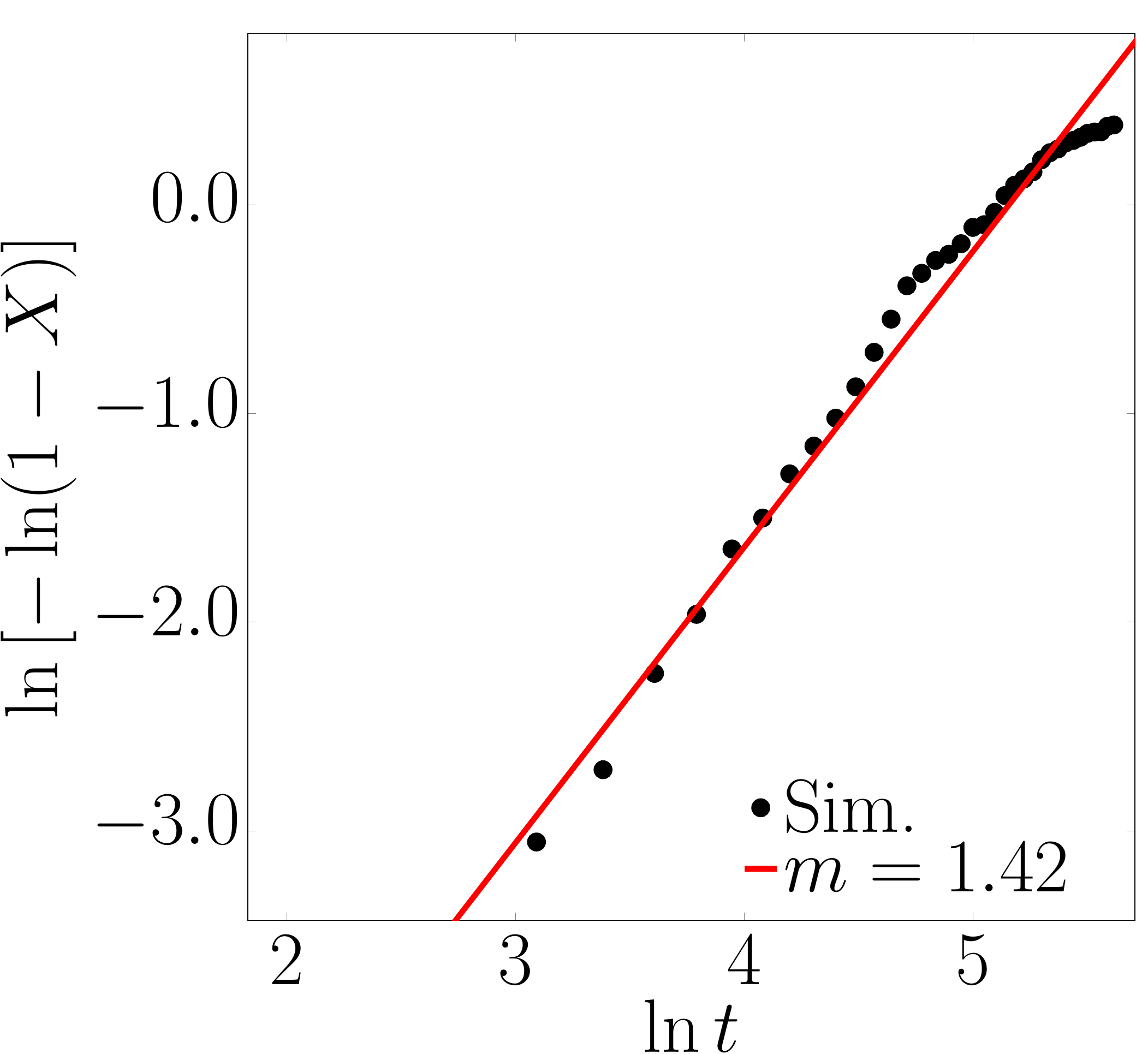}}
\hspace{10pt}
\subfigure[]{\label{fig:AvramiFit}\includegraphics[width=0.42\textwidth]{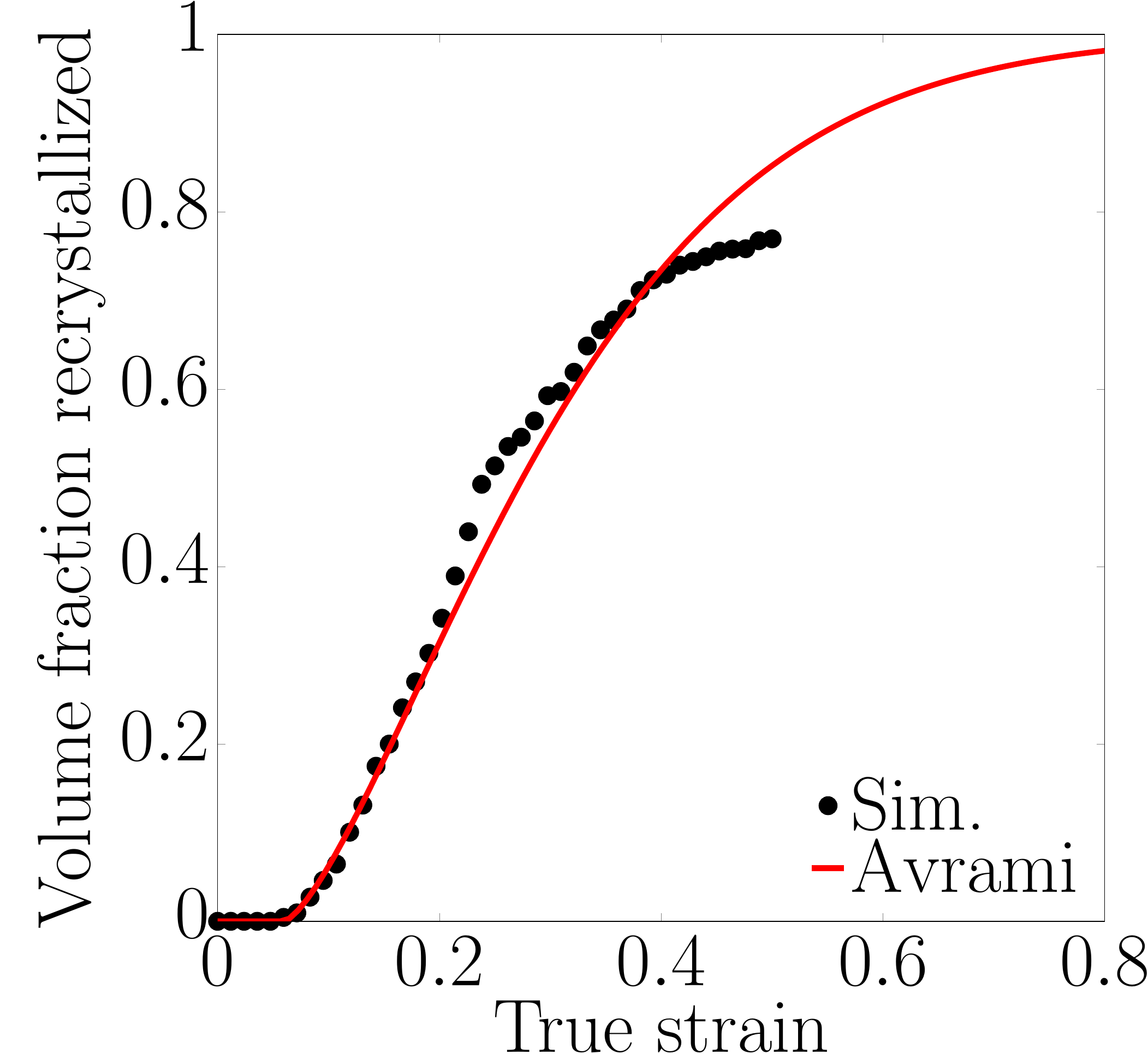}}
\caption{Avrami analysis of simulated DRX kinetics: (a) the determination of Avrami exponent $m$ and
(b) the comparison of between simulated and fitted volume fraction recrystallized.}
\label{fig:AvramiAnalysis}
\end{center}
\end{figure*}

For static recrystallization, the kinetics can be analyzed using the Avrami equation:
\begin{equation}
X\;=\;1-\exp\left[-Bt^m\right],
\label{eq:Avrami}
\end{equation}
where $X$ is the volume fraction recrystallized, $t$ is time, $m$ is a constant called Avrami exponent, and $B$ is also
a constant. During DRX if
the strain rate $\dot{\varepsilon}$ is constant, the time in Eq. (\ref{eq:Avrami}) is usually replaced by
$t=(\varepsilon-\varepsilon_c)/\dot{\varepsilon}$ and the kinetics can again be analyzed using the Avrami equation \citep{roberts1979dynamic}, which is shown in Fig. \ref{fig:AvramiAnalysis}.
The Avrami exponent $m=1.42$ from our simulation (Fig. \ref{fig:AvramiExpo})
falls in the range of the experimental values $1.0\sim2.4$ revealed
by the data on commercial purity copper \citep{garcia2000hot}.

The slope in Fig. \ref{fig:AvramiExpo} shows some variation, indicating that
the Avrami exponent is decreasing when deformation becomes large. The nonuniform Avrami exponent has
also been observed in static recrystallization, where the Avrami exponent decreases
as the recrystallization proceeds. This is attributed to the lack of uniformity of the
stored energy \citep{rollett1989computer}, which is also the case in DRX.
The Avrami analysis (Fig. \ref{fig:AvramiFit}) indicates that the completion of DRX is $\sim80\%$, \hl{at which level
the steady-state flow has just been reached according to the experimental stress-strain curve} of \cite{wusatowska2002nucleation}.

The estimated mean grain size $D_{\rm rex}$ during the compression
(Fig. \ref{fig:GrainSizeComp}) shows grain refinement of
$D_{\rm rex}/D_0=0.5$ ($D_0$ is the initial
mean grain size) up to $50\%$ strain, which is consistent with the common
empirical criterion \citep{sakai1984overview}
that a single stress peak DRX usually corresponds to $2D_{\rm rex}\leq D_0$.
The corresponding experiment data \citep{wusatowska2002nucleation} shows
$D_{\rm rex}/D_0=0.2$,
measured after a strain of $\sim1.3$ when the steady-state has been reached.
It is difficult to ``extrapolate'' current simulation results
(Figs. \ref{fig:NumGrain} and \ref{fig:GrainSizeComp}) and compare with this experimental
result. \hl{Nevertheless, the initially rapid grain refinement has shown a decrease in the refinement rate.
In our future work, the replacement of FFT-EVP with a finite-strain framework will eventually allow us to
approach the steady grain size revealed by many DRX experiments} \citep{sah1974grain,blaz1983effect,sakai1984overview}.

\subsection[]{\hl{Softening during DRX}}
\label{sec:softening}
\begin{figure*}[htb!]
\begin{center}
\subfigure[]{\label{fig:SigVM-before1stDRX}\includegraphics[width=0.34\textwidth]{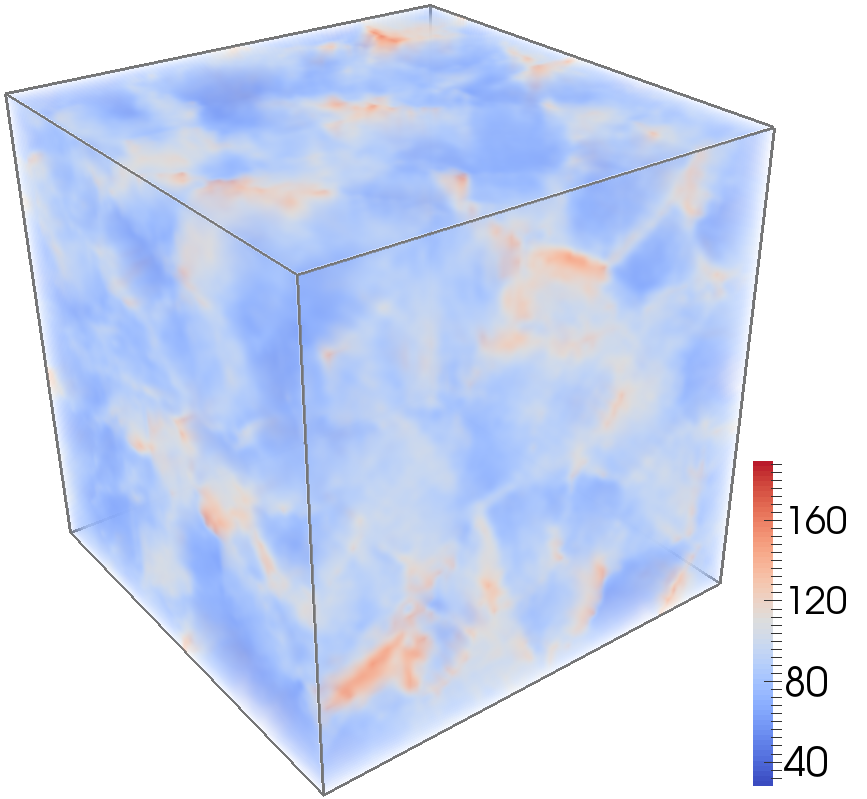}}
\hspace{22pt}
\subfigure[]{\label{fig:DiffSigVM-1stDRX}\includegraphics[width=0.34\textwidth]{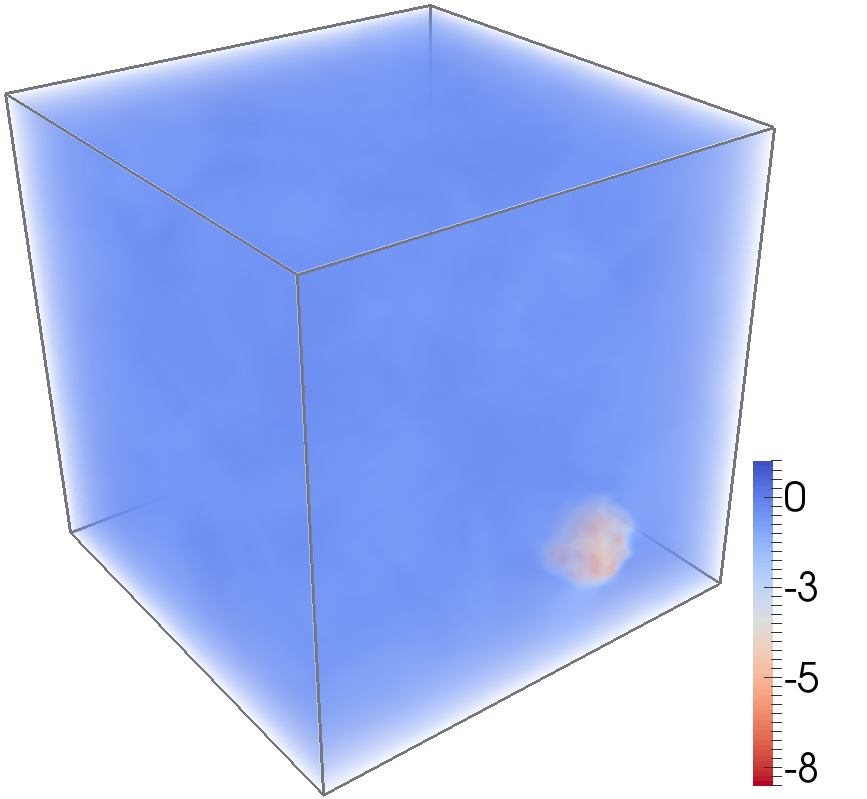}}
\caption{Spatial distribution of (a) von Mises stress prior the nucleation of the first DRX
grain and (b) the resulted change of von Mises stress upon the corresponding phase-field simulation (colorbar unit: MPa).}
\label{fig:1stDRX}
\end{center}
\end{figure*}

While the analogy between
the formation of ``dislocation-free'' grains and phase transformation
has been adopted to intuitively explain the softening of DRX,
it has to be noted that DRX is essentially related to the formation and migration of grain
boundaries, which are extended defects rather than phases, and extended defects are strongly driven by
stress rather than thermodynamics \citep{zhao2013heterogeneously,zhao2014extended}.
At the length scale of discrete extended defects, the RVE-average stress is expected to
exhibit successive drops due to the repeated relaxation of local stresses, which, however, may easily be
averaged out from further coarse-graining at a larger length scale. For instance, while dynamic
recovery due to dislocation annihilation can certainly lead to softening,
the effect is usually balanced by the hardening due to dislocation entanglement occurring
at the same scale (as shown in Eq. (\ref{eq:SSDEvolution})), leaving the
stress-strain curve approaching monotonically a plateau \citep{rollett2004recrystallization}
without showing any macroscopic softening, as shown in the standalone FFT-CP simulation in Figs. \ref{fig:SSs}
and \ref{fig:HRs}.
As a result, softening (negative tangent modulus) during DRX must be attributed to dislocation population evolution due to stress relaxation at a much larger scale than that of discrete dislocation motions/interactions;
which is limited to the mean free path of dislocation slip ($\sim 1/\sqrt{\rho_{\rm tot}}$) and/or the interaction
length of dislocation annihilation (such as climb).
To this end, the  stress field prior to the nucleation of
the very first DRX grain (Fig. \ref{fig:Growth-1stDRX})
is plotted in Fig. \ref{fig:SigVM-before1stDRX}.
Compared with the corresponding grain structure in Fig. \ref{fig:gID-initial}, it is obvious that stress
concentrations
mainly occur at the grain boundaries and junctions, supporting the
GB bulging mechanism adopted in our simulation. (The dislocation density field has a similar distribution.)
The change of von Mises stress after the growth of the new grain, shown in Fig. \ref{fig:DiffSigVM-1stDRX},
indicates that the most significant changes ($>5$ MPa) are localized within
a small region that corresponds exactly to the region of the new grain as shown in Fig. \ref{fig:Growth-1stDRX}.
(Note that Figs. \ref{fig:Growth-1stDRX} and \ref{fig:1stDRX} have the same viewpoint.)
In addition, the maximum decrease of local stress in Fig. \ref{fig:DiffSigVM-1stDRX} is $\sim9$ MPa, which is
at the level of the overall macroscopic stress drop as shown in Fig. \ref{fig:SSs}. 
Similar results are obtained
when analyzing the stress relaxation upon DRX event at the quadruple junction shown in Fig. \ref{fig:Growth-2ndDRX}.
This suggests that enforcing strain rate compatibility immediately after DRX nucleation can effectively redistribute the equilibrium internal stress field to maintain
concordance with the evolved grain structure. The redistributed internal stress can then influence the subsequent
dislocation density (both SSD and GND) evolution through the mesoscopic constitutive laws (Eqs. (\ref{eq:SSDEvolution})
and (\ref{eq:GNDEvolution})) and eventually lead to the macroscopic softening on the stress-strain curve.
The exhibited macroscopic
instability is a dynamic net result developed by a series of successive discrete DRX events
subject to continuous loading conditions, which is by no means trivial to perceive based solely on static argument
such as instantaneous dislocation density.
The current
interpretation of DRX emphasizes the importance of the time and length scale associated with the ``dislocation-free'' concept
in the previous DRX description and suggests that internal stress redistribution is responsible for the macroscopic softening.
The redistribution of strain and stress field during DRX nucleation has actually been indicated recently by
\cite{chauve2015strain} using Digital Image Correlation. It is thus of great interest to carry out more detailed analysis
as in Fig. \ref{fig:1stDRX} to yield a statistically more reliable correlation between stress redistribution and DRX; this will
be done in our future work.

\section{Conclusions}

In this paper an integrated full-field modeling scheme that couples plastic deformation with microstructure evolution has
been established, which provides a general framework for investigating thermomechanical processes. The
key feature of the integrated modeling scheme is the selection of microstructure descriptors that will be
used in both constitutive laws of plasticity and the nucleation and growth of new phases/grains.
As the first demonstration, a FFT-based
elasto-viscoplastic model and a phase-field grain boundary migration model were integrated through the combination of
a dislocation-based crystal plasticity model and a statistical DRX nucleation model, with
the density of dislocations (including SSD, GND, and mobile dislocations) being the selected microstructure descriptors
to couple plastic and microstructural evolution.
After calibration solely based on experimental mechanical response of DRX in polycrystalline copper, the model
delivered fruitful full-field microstructural information, which were
used to provide quantitative description and analysis of the DRX process. Quantitative agreement with experiments in terms of the
kinetics and softening during DRX were achieved.

This effort represents several firsts in the simulation of thermomechanical processes. First and foremost,
it is the first successful demonstration of full coupling between crystal plasticity and phase-field simulations using a spectral (FFT) framework. Secondly, this work also represents the first implementation of a dislocation based constitutive theory
within the FFT-CP framework \hl{with applications to polycrystals}, further demonstrating the utility of the FFT method as a general micromechanical modeling platform.
Finally, to the best of the author's knowledge this work represents the first case where
softening has been handled within the FFT-CP as well.

Despite the fact that DRX is a well studied phenomenon, our integrated simulation makes several new predictions
awaiting experimental verification. Namely we predict the evolution of dislocation densities through
the early and intermediate stages of DRX.
In addition, we predict that the dislocation content of newly nucleated grains must increase rapidly due
to the co-deformation of the new grain with its surrounding neighborhood. This suggests that the observed macroscale softening is largely due to the effect of local stress redistribution on the dislocation density, particularly a drop in the statistically stored dislocation content. We further
predict a much earlier onset of DRX than what have been reported or predicted previously due to
the heterogeneous deformation and stochastic nature of DRX nucleation.
Finally we show that DRX grains formed at both grain boundaries
and junctions (e.g., quadruple junctions) \hl{tend to grow in a wedge-like fashion to maintain a triple line
(not necessarily in equilibrium) with the old grains}.
 
While quantitative agreement of our simulations with experiments have been reached,
the current model has some limitations with respect to the specific DRX simulation explored here.
First of all, the small-strain framework of FFT-EVP prevents us from simulating
the complete DRX process to compare with experiments. Secondly,
the special orientation of nucleated DRX grains (e.g., forming the $\Sigma3$
twin boundary with the parent grain \citep{wusatowska2002nucleation})
has not been considered in the current model.
Finally, the dislocation-based constitutive laws employed and modified in the current model can be further improved
towards a more physics-based manner to account for \hl{features such as GND emission/absorption at grain boundaries or
interfaces} \citep{sun2000observations} \hl{and GND saturation} \citep{kysar2010experimental,oztop2013length}.
The future adoption of the finite-strain FFT-CP,  would allow a complete simulation of DRX and more
accurate prediction of the steady state mean grain size. The integration of finite deformation kinematics with the phase-field is also crucial for the simulation of other technologically relevant thermomechanical processes. 

\section{Acknowledgements}
This work was supported by The National Science Foundation under the DMREF program with Grant No. DMR-1435483. Additional support for S.R.N. and Y.W. comes from the National Science Foundation DMREF program Grant no. DMR-1534826.

\appendix
\section{Poisson statistical model for DRX nucleation}\label{appx:StatModelDerivation}
Consider a unit area of GB with a nucleation propensity $k({\bf x})$\footnote{The convention followed in the statistics literature of denoting random variables in bold unfortunately conflicts with the usual use of bold typeface to denote tensorial or vector quantities. It should be clear from context whether a particular variable is a random scalar or a tensor of vector quantity.} (for clarity, we will omit the dependence of $\bf x$), then due to thermal and atomic structural fluctuations, the number of transformation events $N$ will be a random number and expected to increases with increasing $k$. Given the assumptions stated above, we can say that ${\bf N}(k)$ obeys an inhomogeneous Poisson process \citep{niezgoda2014stochastic}. Correspondingly, we introduce the cumulative hazard function $\Lambda(k)$ that encodes the probability of a subcell transformation. The probability of observing ${\bf N}(k)=n$ sub-cell nucleation events is thus given by
\begin{equation}
\label{eq:prob}
P\left({\bf N}(k)=n\right)\;=\;\frac{1}{n!}\exp\left[-\Lambda(k)\right]\Lambda(k)^n
\end{equation}
In order to link Eq. \ref{eq:prob} to a probability of nucleation we need to determine the minimum number of transformations required to form a stable nucleus, $n^*$, and the form of $\Lambda(k)$.  Recall that we have not specified how the division of a FFT
gridpoint into sub-cells can be done, and due to lack of a specific atomistic mechanisms for DRX nucleation it is unclear what physical events $\Lambda(k)$ describes. Clearly there is a connection between the two. If a finer sub-grid is chosen, then $\Lambda(k)$ is expected to have higher absolute values compared to the case of a coarser sub-grid, \hl{with, however,
the actual reaction rate remained the same}. This non-uniqueness actually
allows us to arbitrarily fix the  sub-grid for our convenience such that $n^*=1$ and fit $\Lambda(k)$ so that the resulting nucleation probability matches experimental data. The probability of nucleation can then be given as 
\begin{equation}
P\left({\bf N}(k)\geq n^*\right)\;=\;1-P\left({\bf N}(k)< n^*\right)\;=\;1-\sum_{n=0}^{n^*-1}P\left({\bf N}(k)=n\right),
\label{eq:NucleationRate}
\end{equation}
\begin{equation}
P\left({\bf N}(k)\geq 1\right)\;=\;1-\exp\left(-\Lambda(k)\right),
\label{eq:SimNucleationRate}
\end{equation}
which will serve as the numerical criterion to check if a FFT gridpoint will undergo DRX nucleation. This is the same as the explicit nucleation model of \cite{simmons2000phase} used in phase-field simulation of phase transformations, as well as the twin nucleation model of \cite{niezgoda2014stochastic}. Following the previous studies of \cite{ding2001coupled} and \cite{rollett2004recrystallization}, the cumulative hazard function adopts an exponential form
\begin{equation}
\label{eq:SimLambdaExpression}
\Lambda(k)\;=\;Ck^q\exp\left(-\frac{Q_{\rm DRX}}{k_{\rm B}T}\right)\;=\;\left(\frac{k}{k_c}\right)^q
\end{equation}
where $C$ and $q$ are material constants. $k_c=\left[C^{-1}\exp\left(\tfrac{Q_{\rm DRX}}{k_{\rm B}T}\right)\right]^{{1}/{q}}$ and can be interpreted as the characteristic threshold that the nucleation propensity needs to exceed in order to yield appreciable
nucleation probabilities.
Substituting Eq. (\ref{eq:SimLambdaExpression}) into Eq. (\ref{eq:SimNucleationRate}) leads
to Eq. (\ref{eq:NumCrt}) in Sec. \ref{sec:Statistical_Nucleation}.

\bibliographystyle{model2-names}
\bibliography{MyBib}

\end{document}